\documentclass[a4paper,11pt]{article}
\pdfoutput=1 % if your are submitting a pdflatex (i.e. if you have
\usepackage{jcappub} % for details on the use of the package, please
\usepackage[utf8]{inputenc}
\usepackage[T1]{fontenc}

\usepackage[mathscr]{eucal}
\usepackage[Symbolsmallscale]{upgreek}
\usepackage{xcolor}
\usepackage{slashed}
\usepackage{hyperref}
\usepackage{soul}
\usepackage[normalem]{ulem}

\newcommand{\be}{\begin{equation}}
\newcommand{\ee}{\end{equation}}
\newcommand{\bea}{\begin{eqnarray}}
\newcommand{\eea}{\end{eqnarray}}
\newcommand{\ba}{\begin{align}}
\newcommand{\ea}{\end{align}}
\DeclareMathOperator{\sign}{sign}

\begin{document}
{\hfill INT-PUB-23-053}
\title{New Constraints on Axion-Like Particles from {\it IXPE} Polarization Data for Magnetars}

\author[a]{Ephraim Gau,}
\author[b]{Fazlollah Hajkarim,}
\author[c,d]{Steven P. Harris,}
\author[a]{P. S. Bhupal Dev,}
\author[e]{Jean-Francois Fortin,}
\author[a]{Henric Krawczynski,}
\author[b]{Kuver Sinha}

\affiliation[a]{Department of Physics and McDonnell Center for the Space Sciences, Washington University, St. Louis, MO 63130, USA}

\affiliation[b]{Department of Physics and Astronomy, University of Oklahoma, Norman, OK 73019, USA}

\affiliation[c]{Institute for Nuclear Theory, University of Washington, Seattle, WA 98195, USA}

\affiliation[d]{Center for the Exploration of Energy and Matter and Department of Physics,
Indiana University, Bloomington, IN 47405, USA}

\affiliation[e]{D{\'e}partement de Physique, de G{\'e}nie Physique et d'Optique,
Universit{\'e} Laval, Qu{\'e}bec, QC G1V 0A6, Canada}

\emailAdd{ephraimgau@wustl.edu}
\emailAdd{fazlollah.hajkarim@ou.edu}
\emailAdd{stharr@iu.edu}
\emailAdd{bdev@wustl.edu}
\emailAdd{jean-francois.fortin@phy.ulaval.ca}
\emailAdd{krawcz@wustl.edu}
\emailAdd{kuver.sinha@ou.edu}

\abstract
{
We derive new constraints on axion-like particles (ALPs) using precision $X$-ray polarization studies of magnetars. Specifically, we use the first detection of polarized $X$-rays from the magnetars 4U 0142+61 and 1RXS J170849.0-400910 by the {\it Imaging $X$-ray Polarimetry Explorer (IXPE)} to place bounds on the product of the ALP-photon and ALP-nucleon couplings, $g_{a\gamma}g_{aN}$, with $g_{aN}$ being responsible for ALP production in the core of the magnetar and $g_{a\gamma}$ controlling the ALP-photon conversion probability in the magnetosphere. These bounds are most sensitive to the magnetar core temperature, and we use two benchmark values of $1\times 10^8$ K and $5\times 10^8$ K to derive our constraints. For the latter choice, our bounds are competitive with the existing bounds on the coupling product coming from a combination of CAST (for $g_{a\gamma}$) and SN1987A (for $g_{aN}$). We advocate for more precise and extensive observational campaigns in the higher end of the 2~--~8~keV spectral window, where ALP-induced polarization is the strongest. We further advocate for hard $X$-ray polarization studies of young, hot, near-Earth magnetars with strong magnetic fields.
}

\maketitle

%%%%%%%%%%%%%%%%%%%%%%%%%%%%%%%%%%%%%%%%%%%%%%%%%%%%%%%%%%%%%%%%%%

\section{Introduction}
\label{sec:introduction}

Axion-like particles (ALPs) are one of the most well-motivated beyond-the-Standard-Model candidates, thus prompting a wide variety of laboratory, astrophysical, and cosmological searches~\cite{Graham:2015ouw, Irastorza:2018dyq, Choi:2020rgn}. An increasingly interesting astrophysical probe is examining their effects on the polarization of light. This effect arises from the fact that ALPs propagating in a background magnetic field will be converted to photons polarized parallel to the magnetic field, and this conversion would alter the polarization measurements away from the values predicted using solely regular astrophysical processes. Applications of this physics include birefringence with the Cosmic Microwave Background~\cite{Carroll:1989vb, Harari:1992ea, Carroll:1998zi,Lue:1998mq, Finelli:2008jv, Fedderke:2019ajk, BICEPKeck:2021sbt, Diego-Palazuelos:2022dsq, SPT-3G:2022ods, POLARBEAR:2023ric}, black holes of various magnitudes (including stellar mass black holes and active galactic nuclei)~\cite{Arvanitaki:2010sy, Plascencia:2017kca, Chen:2019fsq,Gussmann:2021mjj, Yuan:2020xui, Chen:2021lvo, Chen:2022oad, Shakeri:2022usk, Gan:2023swl, Marsh:2017yvc, Ivanov:2018byi}, and pulsars~\cite{Liu:2019brz, Poddar:2020qft}. 
However, these birefringence signals are only relevant for ultralight ALPs with masses $\lesssim 10^{-11}$ eV. In this paper, we point out a new window of opportunity using precision $X$-ray polarization studies of magnetars, which can probe ALPs up to about $10^{-6}$ eV, complementary to the haloscope searches~\cite{ADMX:2021nhd}, as well as previous polarimetry studies at other wavelengths with magnetic white dwarfs~\cite{Gill:2011yp, Dessert:2022yqq}. 

The {\it Imaging X-Ray Polarimetry Explorer (IXPE)}, launched on December 9, 2021, is the first dedicated satellite mission to investigate the $X$-ray polarization of cosmic sources in fifty years~\cite{2022JATIS...8b6002W}. {\it IXPE} obtained spectral and linear polarization data from the magnetars 4U 0142+61 (from January to February 2022) and 1RXS J170849.0-400910 (from September to October 2022). Among other results, these observations revealed a 90-degree swing of the polarization direction, or angle, when going from lower (2 keV) to higher (8 keV) energies for the first magnetar~\cite{2022Sci...378..646T}. A very high linear polarization degree, above 50\% in the higher-energy bins, was detected for the second magnetar~\cite{2023ApJ...944L..27Z}. 

Given the extremely strong magnetic fields ($\sim 10^{14}$ G) of magnetars---surpassing the magnitude of the QED critical field---the magnetospheres of these stars provide some of the best environments to constrain the conversion of ALPs to photons. If such conversion occurs, it will change the $X$-ray energy and polarization spectra reaching the observer. Emission from magnetars therefore holds out the promise of yielding \textit{precision} constraints on ALP couplings: the localized conversion in a relatively well-defined dipolar field, in this case, should be contrasted with ALP-photon conversion across vast (inter)galactic distances, which are prone to far greater uncertainties, especially concerning the magnetic field direction and strength across such vast regions (we refer to ~\cite{Gong:2016zsb, Csaki:2001yk, Mirizzi:2006zy, Conlon:2018iwn, Schallmoser:2021sba} for a sample of this vast literature). Moreover, ALP-photon conversion at $X$-ray energies is free from the uncertainties of plasma effects that are relevant at radio energies~\cite{Lai:2006af, Tjemsland:2023vvc, Prabhu:2023cgb}. Indeed, the main source of uncertainty at $X$-ray energies is the rate of ALP production in the core, which scales strongly with the core temperature \cite{Fortin:2021sst}.  As the core temperature of the magnetars is uncertain (magnetars have anomalously high surface temperatures \cite{Beloborodov:2016mmx}), the standard practice is to show the final constraints for several benchmark temperatures (\textit{e.g.},\ Refs.~\cite{Fortin:2021sst,Lloyd:2020vzs,Maruyama:2017xzl}). Thus, we do so for reasonable core temperatures of $1\times10^8$ K and $5\times10^8$ K. 

The underlying theoretical formalism of extracting bounds on ALP couplings from the intensity and polarization of $X$-ray emission from magnetars
%, following these broad contours, 
has been established by a subset of the present authors~\cite{Fortin:2018ehg, Fortin:2018aom, Fortin:2021sst, Fortin:2021cog, Fortin:2023jlg}. The purpose of this paper is to demonstrate the application of that procedure, extracting the first constraints using polarization data from \textit{IXPE} observations of 4U 0142+61 and 1RXS J170849.0-400910 on the product of the ALP-nucleon and ALP-photon couplings. 

The overall physical picture is the following: ALPs produced in the magnetar core by nucleon bremsstrahlung stream out to the magnetosphere, at which point they convert to $X$-ray photons, polarized along the direction parallel to the magnetic field at the point of conversion. To obtain the theoretical ALP-induced polarization signal, one must thus compute the following:
\begin{itemize}
    
    \item {\it ALP luminosity}: ALPs are produced in the magnetar core by nucleon bremsstrahlung processes $N+N'\rightarrow N+N'+a$ ($N$ and $N'$ are neutrons or protons). The local production rate is calculated in the strongly degenerate limit, taking into account the superconductivity of the protons, which results in a density-dependent gap in their energy spectrum. Several different nuclear equations of state (EoSs) and models for the proton pairing are used to obtain a range of production rates. The local production rate is integrated over spherically symmetric neutron star profiles in order to obtain the ALP luminosity. In constructing the profiles, a range of magnetar masses is considered, as the mass of the two magnetars studied here is unknown. Much of the formalism is the same as in Ref.~\cite{Fortin:2021sst}, though in this work we incorporate general relativistic effects.  
    %that should have been included in Ref.~\cite{Fortin:2021sst} (see footnote \ref{footnote:improve}.).

    \item {\it Conversion Probability and Stokes Parameters:} The calculation of the conversion probability in the magnetosphere, and the resultant expressions for how the Stokes parameters $I$ and $Q$ change after the contribution of ALPs, involves solving a system of coupled differential equations describing the propagation of ALPs in a dipolar magnetic field. Although the full numerical solutions have been used to obtain the constraints in this work, useful semi-analytic solutions of these equations were obtained in Ref.~\cite{Fortin:2018aom}, presented in greater detail in Ref.~\cite{Fortin:2021sst}, and finally extended to include semi-analytic expressions for all four of the Stokes parameters in Ref.~\cite{Fortin:2023jlg}.

\end{itemize}
At this juncture, an important point about the astrophysical `background' should be clarified. 
The astrophysics of all aspects of magnetar $X$-ray emission---including the production and spectropolarimetric form of the surface thermal emission and magnetospheric scattered emission---are subjects of ongoing research (see Refs.~\cite{2011ApJ...730..131F, 2014MNRAS.438.1686T, 2019MNRAS.483..599G, 2020MNRAS.492.5057T, 2022A&A...658A.161K, 2022MNRAS.514.5024C, 2023PNAS..12016534L} for just a small subset of the work). Indeed, the \textit{IXPE} magnetar campaigns also present their own astrophysical modeling of the polarization of the observed emission~\cite{2022Sci...378..646T, 2023ApJ...944L..27Z}.
However, our work is interested in ALPs, which constitute new physics over and above these standard astrophysical effects. Our goal is to obtain reliable upper bounds on ALP couplings by demanding that any new physics contributions to the polarization do not differ from the observational results by more than the stated uncertainty. Thus, we will assume that purely standard-model astrophysics models completely reproduce the observed polarization (and in instances where such concordance does not yet exist, we do \textit{not} claim an anomaly or a resolution with ALP physics). Additionally, as with all ALP studies coming from astrophysics, our constraints depend on how conservative we are with respect to our choice of astrophysical parameters that determine the production of ALPs from the core. For context, we will depict and compare our results to constraints coming from the CAST helioscope experiment~\cite{CAST:2017uph} and from SN1987A cooling arguments~\cite{Raffelt:2006cw, Giannotti:2017hny}, as well as comment on recent astrophysical constraints obtained by other groups~\cite{Beznogov:2018fda, Dolan:2022kul, Noordhuis:2022ljw}. 

We organize the rest of our paper as follows. We summarize relevant concepts and equations from the theoretical polarization formalism in Section~\ref{sec:poltheory}. We next describe the {\it IXPE} experimental context and observational data in Section~\ref{sec:data}. We then describe our method for obtaining the ALP conversion constraints (on the product of ALP-photon and ALP-nucleon couplings, $g_{a\gamma}g_{aN}$) and show our results in Section~\ref{sec:results}. We finally give our conclusions and implications for future studies in Section~\ref{sec:conclusion}.

%%%%%%%%%%%%%%%%%%%%%%%%%%%%%%%%%%%%%%%%%%%%%%%%%%%%%%%%%%%%%%%%%

\section{ALP production and polarization formalism}
\label{sec:poltheory}

We consider ALPs that have a coupling to nucleons as well as to photons. The relevant parts of the Lagrangian are thus given by 
\be
\mathcal{L}\supset-\frac{g_{a\gamma}}{4}aF_{\mu\nu}\tilde{F}^{\mu\nu}+g_{aN}(\partial_{\mu}a)\bar{N}\gamma^{\mu}\gamma_5N\, .
\ee
Here, $a$ denotes the ALP field, $F^{\mu\nu}$ is the electromagnetic field strength tensor, and $N$ denotes the nucleon field. The coupling constants %$g\equiv g_{a\gamma}$ 
$g_{a\gamma}$
and $g_{aN}$ have mass dimension $-1$. We treat them as independent parameters, although in specific models, they can be related to each other by ${\cal O}(1)$ coefficients~\cite{Kim:1979if,Shifman:1979if, Zhitnitsky:1980tq,Dine:1981rt}. In principle, we could have only chosen to work with the $g_{a\gamma}$ coupling, in which case the ALP production would be governed by the Primakoff and photon coalescence processes. However, in presence of the $g_{aN}$ coupling, the production rate is enhanced, being dominated by nucleon bremsstrahlung, which we assume to be the case for this study.

\subsection{ALP production}\label{SProd}
Neutron stars, including magnetars, consist of a thin outer crust of nuclei, arranged in a Coulomb lattice, enclosing an approximately 10-km-radius core of uniform dense matter. Around nuclear saturation density ($n_0\equiv 0.16 \text{ fm}^{-3}$), the matter consists of a charge-neutral neutron-proton-electron fluid. The center of the neutron star might contain matter of density several times that of $n_0$, where additional degrees of freedom (free quarks, hyperons, etc.) might exist. In this work, for simplicity, we will assume that the entire neutron star core consists of a uniform, charge-neutral, and beta-equilibrated fluid of neutrons, protons, and electrons (termed `$npe$ matter'). We will focus on the core, as opposed to the crust, since the dominant contribution to ALP emission comes from the core~\cite{Sedrakian:2018kdm, Fortin:2021cog}.  
    
Even with the assumption of uniform $npe$ matter, there are many uncertainties concerning the nature of dense matter. The nuclear EoS is uncertain above saturation density. Of relevance to the ALP production rate, the proton fraction $x_p \equiv n_p/n_B$ (where $n_p$ is the proton number density and $n_B$ is the baryon number density) has various predicted behaviors as density increases~\cite{Tsang:2023vhh}. In addition, we expect proton Cooper pairing in some density range in the neutron star interior, but the size of the gap produced (or equivalently, the critical temperature above which the pairing ceases) is poorly constrained~\cite{Sedrakian:2018ydt}. Finally, the masses of the magnetars observed is unknown, which means that we do not know the central density reached in their interior. To account for these uncertainties, we  produce a band of predicted ALP production spectra (described further below), representing the range of possible production rates given the above uncertainties.

To examine the impact of these uncertainties, we must first discuss the three nucleon bremsstrahlung processes primarily responsible for producing the ALPs which couple to nucleons:
\begin{subequations}
\begin{align}
    n+n&\rightarrow n+n+a\label{eq:rxn1}\\
    n+p&\rightarrow n+p+a\label{eq:rxn2}\\
    p+p&\rightarrow p+p+a.\label{eq:rxn3}
\end{align}
\end{subequations}
ALP production is reviewed in Refs.~\cite{Raffelt:1996wa,Fortin:2021cog}.\footnote{In superfluid/superconducting nuclear matter, ALPs can also be produced by Cooper pair formation processes, as described in Ref.~\cite{Fortin:2021cog}, but we will neglect those ALPs, which are of too high energy to convert to keV photons.} The two magnetars studied here are fairly young, with the older one having a characteristic age of only 68 kyr (see Table~\ref{tab:magValues}). Neutron star cooling calculations that account for magnetic fields and superfluidity~\cite{2012MNRAS.422.2632H,Ho:2017bia,Potekhin:2017ufy} indicate that such stars likely still have a core temperature above $10^{8} \text{ K}$.  Based on various constraints on superfluidity in neutron stars,\footnote{An analysis in Ref.~\cite{Beznogov:2018fda} determines that the supernova remnant HESS J1731-347 must have matter with proton triplet pairing with a critical temperature above $4\times 10^9\text{ K}$ in much of the core, while neutron triplet pairing must have a critical temperature less than $3\times 10^8\text{ K}$ in the entire core. For a review of superfluidity in neutron stars, see Ref.~\cite{Sedrakian:2018ydt}.} 
the matter inside the magnetars is likely above the neutron triplet pairing critical temperature, and thus only protons experience pairing in some or all of the magnetar core. In terms of pairing mechanisms, we therefore only consider proton singlet pairing in our calculations of ALP production, and thus the rate of $n+n\rightarrow n+n+a$ can be calculated without considering superfluid effects.  We discuss this rate calculation first.

The reaction $n\rightarrow n+a$ is kinematically forbidden, thus requiring a spectator neutron to interact with the initial neutron to satisfy the kinematics. This leads to the reaction $n+n\rightarrow n+n+a$. The strong interaction between the two neutrons is modeled by the one-pion-exchange (OPE) interaction (see the discussion of OPE in Ref.~\cite{Fortin:2021cog}), though with a correction factor of $C_{\pi}=1/4$ as suggested in Ref.~\cite{Hanhart:2000ae}. The matrix element of the reaction and the details of the emissivity calculation are given in Ref.~\cite{Fortin:2021sst}. The ALP emissivity spectrum from the reaction $n+n\rightarrow n+n+a$, which is a local quantity defined for a single fluid element in the magnetar, is given by
\begin{equation}
    \frac{\mathop{dQ^0_{nn}}}{\mathop{d\omega}_a} = \frac{1}{36\pi^7}C_{\pi} \left(\frac{m_N}{m_{\pi}}\right)^4 f^4 g_{aN}^2p_{Fn} F(c)\omega_a^3\frac{\omega_a^2+4\pi^2T^2}{e^{\omega_a/T}-1}.
    \label{eq:axion_nn_emissivity}
\end{equation}
Here $m_N$ is the nucleon mass, $m_\pi$ is the pion mass, $f\approx1$ is the pion-nucleon coupling constant~\cite{Friman:1979ecl}, $\omega_a$ is the ALP energy, $T$ is the local temperature, $p_{Fn}$ is the neutron Fermi momentum, $c \equiv m_{\pi}/(2p_{Fn})$, and the function $F(c)$ is given in Eq.~(2.5) of Ref.~\cite{Fortin:2021sst}. The superscript zero in $\mathop{dQ^0_{nn}}/\mathop{d\omega}_a$ denotes that the neutrons are unpaired.  

The production processes involving protons must be discussed separately, because the Cooper pairing of protons opens up a gap $\Delta(T,n_B)$ in the energy spectrum at the Fermi energy, exponentially suppressing the ALP production rates.  The ALP emissivity spectrum from the $n+p\rightarrow n+p+a$ reaction is 
\begin{equation}
    \frac{\mathop{dQ^S_{np}}}{\mathop{d\omega}_a} = \frac{2}{3\pi^7}C_{\pi}\left(\frac{m_N}{m_{\pi}}\right)^4f^4g_{aN}^2p_{Fp}G(c,d)T^3\omega_a^2 I^S_{np}(y=\omega_a/T,n_B,T),
    \label{eq:dQSnpdx}
\end{equation}
where
\begin{equation}
    I^S_{np}(y,n_B,T)=\int_{-\infty}^{\infty}\mathop{dx_2}\mathop{dx_4}\frac{z_4+y-z_2}{(1+e^{z_2})(1+e^{-z_4})(e^{z_4+y-z_2}-1)},
    \label{eq:2.5}
\end{equation}
where $z_i$ is defined as
\begin{equation}
    z_i = \sign(x_i)\sqrt{x_i^2+(\Delta(T,n_B)/T)^2}.
    \label{eq:z}
\end{equation}
The function $G(c,d)$, where $d\equiv m_{\pi}/(2p_{Fp})$ in terms of the proton Fermi momentum $p_{Fp}$, is given in Appendix B of Ref.~\cite{Fortin:2021sst}. 
The superscript $S$ in $\mathop{dQ^S_{np}}/\mathop{d\omega}_a$ and $I^S_{np}$ denotes that the protons are superconducting.

The ALP emissivity spectrum due to $p+p\rightarrow p+p+a$ can be written as
\begin{equation}
    \frac{\mathop{dQ^S_{pp}}}{\mathop{d\omega}_a} = \frac{2}{3\pi^7}C_{\pi}\left(\frac{m_N}{m_{\pi}}\right)^4f^4g_{aN}^2p_{Fp}F(d)T^3\omega_a^2 I^S_{pp}(y=\omega_a/T,n_B,T),
    \label{eq:dQSnpdx1}
\end{equation}
where
\begin{equation}
    I^S_{pp}(x,n_B,T)=\int_{-\infty}^{\infty}\mathop{dx_1}\mathop{dx_2}\mathop{dx_3}\mathop{dx_4}\frac{\delta(z_1+z_2-z_3-z_4-y)}{(1+e^{z_1})(1+e^{z_2})(1+e^{-z_3})(1+e^{-z_4})}.
\end{equation}
The detailed derivations of these emissivity expressions are given in Refs.~\cite{Fortin:2021sst,Yakovlev:1999sk}. To prepare for integrating these emissivities over the neutron star interior, we discuss our specific treatment of $npe$ matter and the proton superconductivity.  

Calculations of the superfluid gap in the proton spectrum must be done through models of the nuclear interaction. We use the CCDK model as parameterized in Ref.~\cite{Ho:2014pta} (see Eq.~(2) and Table~2 in their paper), which predicts the size of the density-dependent gap $\Delta(n_B)$ at zero temperature. The size of the gap at finite temperature can be calculated with the BCS gap equation as described in Ref.~\cite{Yakovlev:1999sk} (or see the discussion in Ref.~\cite{Fortin:2021sst}). Then, a finite $\Delta$ suppresses the integral $I^S$, compared to the case when $\Delta=0$.

We model $npe$ matter with a variety of relativistic mean field (RMF) theories.  In an RMF, the strong force between nucleons is modeled by the exchange of mesons, including the scalar $\sigma$ and the vectors $\omega$ and $\rho$.  Relativistic mean field theories differ in the types of nonlinear couplings the mesons are allowed to have, as well as the meson/nucleon and meson/meson coupling strengths, which are fit to properties of nuclear matter. For comprehensive reviews of RMFs, see Refs.~\cite{Dutra:2014qga,Glendenning:1997wn}. We will use the IUF~\cite{Fattoyev:2010mx}, BSR12~\cite{Dhiman:2007ck}, NL$\rho$~\cite{Liu:2001iz}, HC~\cite{Bunta:2003fm}, and FSU Garnet~\cite{Utama:2016tcl} EoSs. They are each roughly consistent with an array of nuclear and astrophysical constraints (see Ref.~\cite{Nandi:2018ami}).

Once an EoS has been specified, and a central density has been chosen, the Tolman– Oppenheimer–Volkoff (TOV) equations can be solved to obtain the enclosed mass $m(r)$ and pressure $P(r)$ profiles, from which the total mass $M$, the radius $R$, and quantities like the Fermi momenta $p_{F,i}(r)$ can be obtained~\cite{Tolman:1939jz,Oppenheimer:1939ne,Glendenning:1997wn}.  

Ultimately, we seek the differential ALP luminosity, obtained by integrating the total emissivity over the magnetar. As the ALPs are emitted and climb out of the gravitational potential well of the magnetar, their energy is redshifted. ALP conversion to photons is most probable at a distance of $r\approx 1000\,R$ (where $R\simeq 12$ km is the equatorial radius of the magnetar)~\cite{Fortin:2021sst}, essentially infinitely far away from the perspective of gravitational redshift. Thus, we calculate the ALP differential luminosity observed at infinity,
\begin{equation}
    \dfrac{\mathop{dL^{\infty}}}{\mathop{d\omega_a^{\infty}}}=4\pi\int_0^{R_{\text{crust}}}\mathop{dr}\dfrac{r^2e^{\phi(r)}}{\sqrt{1-\frac{2m(r)G}{r}}}\dfrac{\mathop{dQ_{\text{total}}}}{\mathop{d\omega_a}},
\end{equation}
where $G$ is Newton's constant, $e^{\phi(r)} = \sqrt{-g_{00}}$ is the gravitational redshift factor~\cite{1977ApJ...212..825T}, $R_{\rm crust}$ is the core-crust transition location from the center ($R_{\rm crust}\approx 11$ km for a $1.4M_{\odot}$ IUF neutron star \cite{Fortin:2021sst}), and the total emissivity spectrum $\mathop{dQ_{\text{total}}}/\mathop{d\omega_a}$ is the sum of the three contributions from Eqs.~\eqref{eq:rxn1}-\eqref{eq:rxn3}. Noting that $\omega^{\infty}=\omega_{\text{local}}\,e^{\phi}$~\cite{Potekhin:2015qsa}, the differential, redshifted luminosity for $n+n\rightarrow n+n+a$, for example, would be given by the integral
\begin{align}
     \dfrac{\mathop{dL^{\infty}_{nn}}}{\mathop{d\omega_a^{\infty}}} &= \dfrac{1}{9\pi^6}C_{\pi}\left(\frac{m_N}{m_{\pi}}\right)^4f^4g_{aN}^2\dfrac{\omega_{a}^{\infty,3}(\omega_{a}^{\infty,2}+4\pi^2T^{\infty,2})}{e^{\omega_{a}^{\infty}/T^{\infty}}-1}\\
     &\times \int_0^{R_{\text{crust}}}\mathop{dr}\dfrac{r^2}{\sqrt{1-\dfrac{2Gm(r)}{r}}}p_{Fn}(r)F[c(r)]e^{-4\phi(r)}.\nonumber
\end{align}
The magnetar is assumed to be isothermal, meaning that its redshifted temperature $T^{\infty}$ is constant throughout the core and can be taken out of the integral~\cite{Yakovlev:2004iq}. The luminosities of the other two bremsstrahlung processes are turned into redshifted expressions in an analogous way.

\begin{figure}
  \centering
  \includegraphics[width=0.45\textwidth]{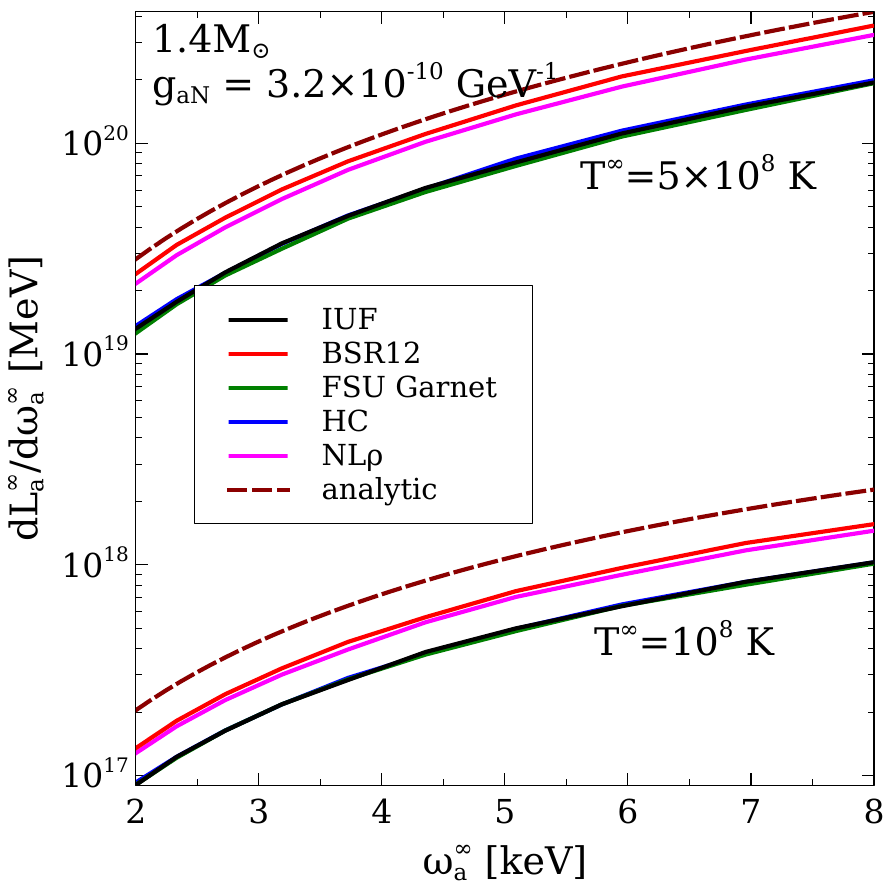}
  \includegraphics[width=0.45\textwidth]{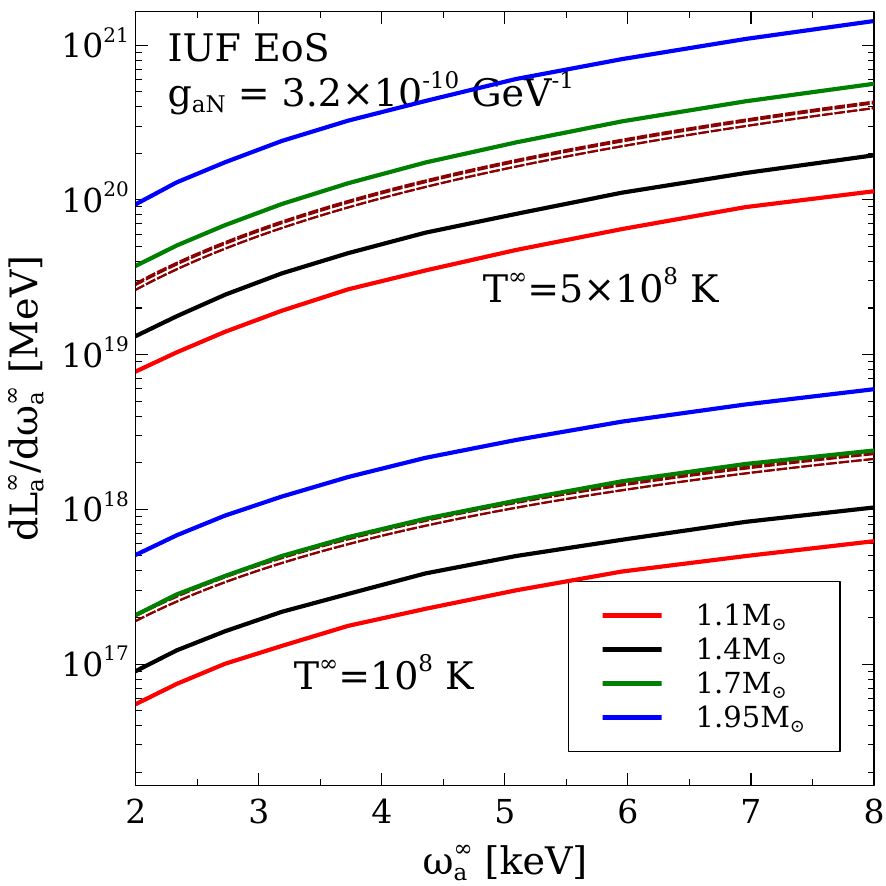}
  \caption{{\it Left panel}: Various ALP spectra found from varying the nuclear EoS, while fixing the magnetar mass at $1.4M_{\odot}$. The upper (lower) set of curves corresponds to a redshifted core temperature $T^\infty=5~(1)\times 10^8$ K. {\it Right panel}: Various ALP spectra for different magnetar masses, fixing the EoS to be IUF.  In both panels, the dashed dark red lines correspond to the analytic estimate of the spectrum given in Eq.~\eqref{prod1}. In the right panel, there is a set of four, essentially overlapping, dashed red lines for each temperature, showing that the analytic approximation of the spectra is essentially independent of the magnetar mass, while the full calculation of the spectra has significant dependence on the magnetar mass. The ALP-nucleon coupling is taken to be the maximum value allowed by SN1987A~\cite{Giannotti:2017hny}, $g_{aN}=3.2\times 10^{-10}\text{ GeV}^{-1}$.}
  \label{fig:ALP_dLdw}
\end{figure}

In the left hand panel of Fig.~\ref{fig:ALP_dLdw}, we plot these results for the ALP spectra for a $1.4M_{\odot}$ magnetar at two different core temperatures, for a range of plausible nuclear EoSs.  The EoSs affect the spectrum through the variation in the predicted proton fraction as well as in the neutron star structure, which affects the redshift. However, changing the EoS has a small overall effect; the boundaries of this variation span only a factor of 2. In the right hand panel, the magnetar mass, which is also unknown, is varied from 1.1 to 1.95 solar masses (the highest value allowed by the IUF EoS, which is fixed in this right panel). The uncertainty in the spectrum due to the magnetar mass variation is one order of magnitude, larger than that due to EoS variation alone.\footnote{Our previous work, Ref.~\cite{Fortin:2021sst}, did not include general relativistic effects in the volume integration, which are particularly important for high-mass neutron stars. Thus, the outer edge of the error band on the ALP constraints might be in actuality slightly more constraining than was depicted in that work, though for that study, as with this study, the unknown core temperature is still by far the greatest uncertainty.\label{footnote:improve}}

In each panel of Fig.~\ref{fig:ALP_dLdw}, a dashed red line displays an analytic approximation for the ALP spectrum: 
\begin{equation} \label{prod1}
    \dfrac{\mathop{dL}}{\mathop{d\omega}_a}=\frac{1}{3^{5/3}\pi^{16/3}}\frac{m_N^{11/3}}{m_{\pi}^4}g_{aN}^2R^3\rho^{1/3}\frac{\omega_a^3(\omega_a^2+4\pi^2T^2)}{e^{\omega_a/T}-1}.
\end{equation}
In this analytic approximation, the neutron star is treated as a constant-density object with mass $M$ and radius\footnote{In this approximation, no attempt is made to distinguish the core (which copiously produces ALPs) and the crust (where ALP production is from a different mechanism, and is weaker)~\cite{Sedrakian:2018kdm,Fortin:2021cog}.} $R$, and thus some average mass density $\rho=3M/(4\pi R^3)$ (typically a few times the nuclear saturation density $\rho_0\approx2.8\times10^{14}\,\text{g}\cdot\text{cm}^{-3}$.)  In Fig.~\ref{fig:ALP_dLdw}, the analytic approximation curve is evaluated for the neutron star masses specified in each panel, with the corresponding radii predicted by the IUF EoS.  The analytic approximation does not account for the nonuniform density distribution, gravitational redshift, nucleon pairing, or beyond-OPE effects, yet (for $T\gtrsim 10^{8}$ K where neutrons remain unpaired), it falls well within the uncertainty band obtained by more sophisticated calculations. 

Also as displayed in the right hand panel of Fig.~\ref{fig:ALP_dLdw}, an even greater variation in the spectrum is obtained due to the choice of core temperature than due to the different EoSs and magnetar masses. Thus, in the below calculations for the constraints, for each core temperature $T^\infty$ ($10^8$ and $5\times 10^8$ K), we take the most conservative and most optimistic set of EoSs and magnetar masses to form an uncertainty band (bounded by the `low luminosity' and `high luminosity' cases displayed in sections below) for the ALP spectrum produced in the core.

\subsection{Calculation of ALP intensity and polarization effect}

In this section, we briefly summarize the results of Ref.~\cite{Fortin:2023jlg}, where the underlying theory was presented for obtaining all four Stokes Parameters in the context of core-produced ALPs converting in the magnetosphere. Readers interested in further details are referred to that paper.

As the emitted ALPs and photons propagate in the magnetosphere, the spatial evolution of the ALP-photon system can be described by a set of coupled equations~\cite{Raffelt:1987im, Lai:2006af}:
\be\label{EqnEE}
i\frac{d}{dx}\left(\begin{array}{c}a\\E_\parallel\\E_\perp\end{array}\right)=\left(\begin{array}{ccc}\omega R+\Delta_aR&\Delta_MR&0\\\Delta_MR&\omega R+\Delta_\parallel R&0\\0&0&\omega R+\Delta_\perp R\end{array}\right)\left(\begin{array}{c}a\\E_\parallel\\E_\perp\end{array}\right)\,\,,
\ee
where 
\be
\Delta_a=-\frac{m_a^2}{2\omega},\qquad\Delta_\parallel=\frac{1}{2}q_\parallel\omega\sin^2\theta,\qquad\Delta_\perp=\frac{1}{2}q_\perp\omega\sin^2\theta,\qquad\Delta_M=\frac{1}{2}g_{a\gamma}B\sin\theta\,\,.
\ee
The photon electric fields parallel and perpendicular to the local magnetic field $B=B_0(R/r)^3$ (in the dipole regime) are denoted by $E_\parallel(x)$ and $E_\perp(x)$, respectively. Characteristic values of the surface equatorial magnetic field $B_0$ (along with other key parameters) for the two magnetars used in our study are given in Table~\ref{tab:magValues}. The variable $x$ is given by $x=r/R$, where $r$ is the distance from the center of the magnetar. The energy of the (axion or photon) particle is $\omega$, and the angle between the dipolar magnetic field and the ALP propagation direction is denoted by $\theta$.
$q_\parallel$ and $q_\perp$ are dimensionless functions, given by
\be
\begin{gathered}
q_\parallel=\frac{7\alpha}{45\pi}b^2\hat{q}_\parallel,\qquad\qquad\hat{q}_\parallel=\frac{1+1.2b}{1+1.33b+0.56b^2},\\
q_\perp=\frac{4\alpha}{45\pi}b^2\hat{q}_\perp,\qquad\qquad\hat{q}_\perp=\frac{1}{1+0.72b^{5/4}+(4/15)b^2},
\end{gathered}
\ee
where $b=B/B_c$, $B_c=m_e^2/e=4.414\times10^{13}\,\text{G}$ is the critical QED magnetic field strength~\cite{Lai:2006af,Raffelt:1987im}, and $\alpha=e^2/(4\pi)\approx1/137$ is the fine structure constant. 

It is convenient to express the different fields on the left hand side of Eq.~\eqref{EqnEE} in terms of the amplitudes $A$ and  phases $\phi$: 
\bea
a(x)&=&A\cos[\chi(x)]e^{-i\phi_a(x)},\nonumber\\
E_\parallel(x)&=&A\sin[\chi(x)]ie^{-i\phi_\parallel(x)},\nonumber\\
E_\perp(x)&=&A_\perp e^{-i\phi_\perp(x)},
\eea
where $A_a=A\cos[\chi(x)]$, $A_\parallel=A\sin[\chi(x)]$, and $A_\perp$ are the amplitudes at position $xR$ of the ALP field, the parallel photon field, and the perpendicular photon field, respectively. The angle $\chi(x)$ parametrizes the mixing of the ALP with the parallel component of the photon. The conservation of probability is then implicitly satisfied~\cite{Fortin:2018ehg}:
\be
\frac{d}{dx}[|a(x)|^2+|E_\parallel(x)|^2]=0\,.
\ee
We can also write the intensities of the various fields as
\bea 
I_a(x)&=&A^2\cos^2[\chi(x)], \nonumber\\
I_\parallel(x)&=&A^2\sin^2[\chi(x)], \nonumber \\
I_\perp(x)&=&A_\perp^2,
\eea
and the conservation of probability thus implies that $I_a(x)+I_\parallel(x)=A^2$ is constant (with $I_\perp(x)=A_\perp^2$ manifestly already constant). 

\begin{table*}
     \begin{center}
     \tabcolsep=0.25cm
	\begin{tabular}{c c c c c} 
		\hline
                Magnetar & Field Strength & Distance & Period & Age \\
            \hline
                4U 0142+61 & $2.6\times10^{14}$ G & $3.6(4)$ kpc & $8.68869249(5)$ s & $68$ kyr \\
                1RXS J170849.0-400910 & $9.4\times10^{14}$ G & $3.8(5)$ kpc & $11.00502461(17)$ s & $9.0$ kyr \\
		\hline
	\end{tabular}
    	\caption{Key parameters for the two magnetars used in this study~\cite{2014ApJS..212....6O}. The magnetic field strength is taken to be a factor of 2 larger than the values quoted in the McGill Magnetar Catalog~\cite{baring}.} 
        \label{tab:magValues}
    \end{center}
\end{table*}

It should be noted that the Stokes parameters $Q$, $U$, and $V$ are generally frame-dependent, as opposed to the intensity $I$ that has been previously used to constrain ALPs with the hard $X$-ray emission ~\cite{Fortin:2018ehg, Fortin:2021sst}. When defined in a frame that is corotating with the magnetar (which we call the `magnetar frame'), the relevant Stokes parameters will be labelled with the superscript `mag'. The corotating frame geometry with a star-centered origin has been discussed in detail in Ref.~\cite{Wadiasingh:2017rcq}. The Stokes parameters, when computed in the frame of the detector, will be labelled with the superscript `det'. 

Then, after solving the coupled differential Eqs.~\eqref{EqnEE} that describe the ALP propagation and account for the ALP-to-photon conversion, we find the equations for the Stokes parameters in the magnetar frame to be~\cite{Fortin:2023jlg}
\be
\begin{gathered}
I_\text{obs}^\text{mag}=I_\text{ast}^\text{mag}+\left[I_a(1)-\frac{I_\text{ast}^\text{mag}-Q_\text{ast}^\text{mag}}{2}\right]P_{a\to\gamma},\\
Q_\text{obs}^\text{mag}=Q_\text{ast}^\text{mag}-\left[I_a(1)-\frac{I_\text{ast}^\text{mag}-Q_\text{ast}^\text{mag}}{2}\right]P_{a\to\gamma},\\
U_\text{obs}^\text{mag}=[U_\text{ast}^\text{mag}\cos(\delta\phi_a)-V_\text{ast}^\text{mag}\sin(\delta\phi_a)]\sqrt{1-P_{a\to\gamma}},\\
V_\text{obs}^\text{mag}=[V_\text{ast}^\text{mag}\cos(\delta\phi_a)+U_\text{ast}^\text{mag}\sin(\delta\phi_a)]\sqrt{1-P_{a\to\gamma}},
\end{gathered}
\ee
which can be simplified to
\be
\begin{gathered} \label{approx}
I_\text{obs}^\text{mag}\approx I_\text{ast}^\text{mag}+I_a(1)P_{a\to\gamma},\\
Q_\text{obs}^\text{mag}\approx Q_\text{ast}^\text{mag}-I_a(1)P_{a\to\gamma},\\
U_\text{obs}^\text{mag}\approx U_\text{ast}^\text{mag},\\
V_\text{obs}^\text{mag}\approx V_\text{ast}^\text{mag},
\end{gathered}
\ee
when $\delta\phi_a=\phi_a(\infty)-\phi_a(1)\ll 1$ and $P_{a\to\gamma}\ll1$ for small ALP-photon coupling. Here, the total observed polarization parameters (with subscript `obs'), including the effect of ALPs, are expressed in terms of the polarization parameters (with subscript `ast') due purely to standard astrophysical effects.  
The ALP contribution is contained in the intensity of ALPs produced from the neutron star surface, $I_a(1)$, and the probability of conversion in the magnetosphere, $P_{a\rightarrow \gamma}$

Since the orientation of the magnetar in the detector frame is not observationally constrained for either of the observed sources, we need expresssions for the observed Stokes parameters in the magnetar frame written in terms of the Stokes parameters in the detector frame:
\be
\begin{gathered} \label{key}
I_\text{obs}^\text{det}\approx I_\text{ast}^\text{det}+I_a(1)P_{a\to\gamma}=(1+p_a)I_\text{ast}^\text{det},\\
Q_\text{obs}^\text{det}\approx Q_\text{ast}^\text{det}-I_a(1)P_{a\to\gamma}\cos(2\Delta\psi)=[q-p_a\cos(2\Delta\psi)]I_\text{ast}^\text{det},\\
U_\text{obs}^\text{det}\approx U_\text{ast}^\text{det}+I_a(1)P_{a\to\gamma}\sin(2\Delta\psi)=[u+p_a\sin(2\Delta\psi)]I_\text{ast}^\text{det},\\
V_\text{obs}^\text{det}\approx\pm V_\text{ast}^\text{det},
\end{gathered}
\ee
where $\Delta\psi$ is the (unconstrained) relative angle between the detector and magnetar frames, and
\be \label{eq:pa}
p_a=\frac{I_a(1)P_{a\to\gamma}}{I_\text{ast}^\text{det}},\qquad q=\frac{Q_\text{ast}^\text{det}}{I_\text{ast}^\text{det}},\qquad u=\frac{U_\text{ast}^\text{det}}{I_\text{ast}^\text{det}}.
\ee
Since the ALP contributions thus come in through $p_a$, we discuss its main two components in turn.

In the regime of interest of the ALP mass and coupling for this study (that is, sufficiently light mass and sufficiently weak coupling), it is convenient for the rest of this section to work in the semi-analytic limit derived in Ref.~\cite{Fortin:2018aom}, such that
\be \label{prob1}
P_{a\to\gamma}\approx\left(\frac{\Delta_{M0}R^3}{r_{a\to\gamma}^2}\right)^2\frac{\Gamma\!\left(\frac{2}{5}\right)^2}{25|\Delta_ar_{a\to\gamma}/5|^\frac{4}{5}},
\ee
where $\Delta_{M0}$ is $\Delta_M$ at the location of the surface, and the conversion radius (\textit{i.e.},\ the radial location where most of the ALP-photon conversion occurs) is
\be \label{prob2}
r_{a\to\gamma}=\left(\frac{7\alpha}{45\pi}\right)^{1/6}\left(\frac{\omega}{m_a}\frac{B_0}{B_c}|\sin\theta|\right)^{1/3}R.
\ee
For the values of the masses and couplings relevant to our study, the probability of conversion satisfies $P_{a\to\gamma}\ll1$, the same small-coupling approximation used to linearize Eqs.~\eqref{approx}. 

Turning to the other component of $p_a$, $I_a$ (the ALP emission intensity in counts per unit time per unit energy) can be given (incorporating the analytic estimate of Eq.~\eqref{prod1} for $dL/d\omega$, in contrast to the full numerical details of Section~\ref{SProd}) by  
\begin{align}
I_a(1) & = \frac{S}{4\pi D^2}\frac{1}{\omega_a^{\infty}}\frac{dL^{\infty}_a}{d\omega^{\infty}_a} \label{eq:Ia} \\
& = \frac{Sf^4}{2^23^{{5}/{3}}\pi^{{19}/{3}}D^2}\frac{m_N^{{11}/{3}}}{m_\pi^4}g_{aN}^2T^4R^3\rho^{{1}/{3}}\frac{y^2(y^2+4\pi^2)}{e^y-1},
\label{eq:Ianalytic}
\end{align}
where $D$ is the distance from the magnetar to Earth (see Table~\ref{tab:magValues}), $S$ is the energy-dependent detector effective area (as shown in Fig.~\ref{fig:detector}), and $y=\omega_a^{\infty}/T^{\infty}=\omega_a/T$. Note that Eq.~\eqref{eq:Ia} is exact, which is used in our numerical calculations, but Eq.~\eqref{eq:Ianalytic} provides an analytic approximation that will be used below for $p_a$. 

Combining $I_a$ and $P_{a\to\gamma}$, then, we arrive at the following result for $p_a$, as derived in Ref.~\cite{Fortin:2023jlg}:
\bea
p_a&=&\frac{\Gamma\!\left(\frac{2}{5}\right)^2}{2^{{24}/{5}}3^{{1}/{15}}5^{{2}/{5}}7^{{4}/{5}}\pi^{{19}/{3}}}\frac{Sf^4m_e^{{16}/{5}}m_N^{{11}/{3}}}{\alpha^{{8}/{5}}m_\pi^4D^2I_\text{ast}^\text{det}}(g_{aN}g_{a\gamma})^2T^{{16}/{5}}R^{{21}/{5}}\rho^{{1}/{3}}B_0^{{2}/{5}}\nonumber \\
&& \qquad\qquad\qquad\qquad\qquad\qquad \times |\sin\theta|^{{2}/{5}}\frac{y^{{6}/{5}}(y^2+4\pi^2)}{e^{y}-1}.
\eea
Since the star's core isotropically produces ALPs, it is possible to average over the angle $\theta$.  To do so, one should technically rewrite $\theta$, the angle between the ALP propagation direction and the local magnetic field, in terms of the viewing angle $\theta_v$, the angle between the ALP propagation direction and the magnetic dipole axis~\cite{Wadiasingh:2017rcq}. The average should then be performed for one complete rotation of the magnetar around its rotation axis. This calculation would thus require knowledge of the inclination angle, sometimes denoted $\alpha_i$, the angle between the rotation axis and the magnetic dipole axis. However, since both the viewing angle and the inclination angle are unknown, and since the average over $\theta$ of $[0,\pi)$ usually differs by a factor of at most two (for most viewing and inclination angles) compared to a proper averaging over a more limited range of $\theta$, we simply proceed with $p_a$ averaged over that full range of $\theta$~\cite{Fortin:2022skl}:
\be
p_a=\frac{5^{{3}/{5}}\Gamma\!\left(\frac{2}{5}\right)^3}{2^{{21}/{5}}3^{{1}/{15}}7^{{4}/{5}}\pi^{{19}/{3}}\Gamma\!\left(\frac{1}{5}\right)^2}\frac{Sf^4m_e^{{16}/{5}}m_N^{{11}/{3}}}{\alpha^{{8}/{5}}m_\pi^4D^2I_\text{ast}^\text{det}}(g_{aN}g_{a\gamma})^2T^{{16}/{5}}R^{{21}/{5}}\rho^{{1}/{3}}B_0^{{2}/{5}}\frac{y^{{6}/{5}}(y^2+4\pi^2)}{e^y-1}.
\ee
Moreover, in order to compare theory with the observational data (reported in finite energy bin widths), we must average $p_a$ over a range of energies, thus obtaining
\begin{align}\label{Eqpa}
p_a&=\frac{5^{{3}/{5}}\Gamma\!\left(\frac{2}{5}\right)^3}{2^{{21}/{5}}3^{{1}/{15}}7^{{4}/{5}}\pi^{{19}/{3}}\Gamma\!\left(\frac{1}{5}\right)^2}\frac{f^4m_e^{{16}/{5}}m_N^{{11}/{3}}}{\alpha^{{8}/{5}}m_\pi^4D^2I_\text{ast}^\text{det}}(g_{aN}g_{a\gamma})^2T^{{21}/{5}}R^{{21}/{5}}\rho^{{1}/{3}}B_0^{{2}/{5}}\nonumber \\
& \qquad \qquad\qquad\qquad\qquad\qquad\qquad\qquad\times \int_{y_i}^{y_f}\frac{dy}{\Delta\omega}\,\frac{S(y)y^{{6}/{5}}(y^2+4\pi^2)}{e^y-1}\\\nonumber
&\approx\left(\frac{5.91\times10^{-4}\,\text{counts}\cdot\text{keV}^{-1}\cdot\text{s}^{-1}}{I_\text{ast}^\text{det}}\right)\left(\frac{1\,\text{keV}}{\Delta\omega}\right)\left(\frac{g_{aN}}{10^{-10}\,\text{GeV}^{-1}}\right)^2\left(\frac{g_{a\gamma}}{10^{-10}\,\text{GeV}^{-1}}\right)^2\left(\frac{\rho}{\rho_0}\right)^{{1}/{3}}\\
&\phantom{=}\times\left(\frac{T}{10^9\,\text{K}}\right)^{{21}/{5}}\left(\frac{R}{10\,\text{km}}\right)^{{21}/{5}}\left(\frac{B_0}{B_c}\right)^{{2}/{5}}\left(\frac{1\, \text{kpc}}{D}\right)^2\int_{y_i}^{y_f}dy\,\left(\frac{S(y)}{100\,\text{cm}^2}\right)\frac{y^{{6}/{5}}(y^2+4\pi^2)}{e^y-1}.\nonumber
\end{align}
Here $\omega\in(\omega_i,\omega_f)$ specifies an energy bin, with $y_i=\omega_i/T$, $y_f=\omega_f/T$, and $\Delta\omega=\omega_f-\omega_i$; the effective area of the detector is kept inside the integral, due to its energy-dependence (as shown in Fig.~\ref{fig:detector}). 

From Eq.~\eqref{Eqpa}, we realize that the equation for $p_a$, and thus the constraints we obtain by comparing the purely-astrophysical Stokes parameters with the ALP-inclusive ($p_a$-dependent) Stokes parameters, is independent of the ALP mass. This is a result of using the small mixing approximations (the same as those used for Eq.~\eqref{approx}), causing us to be in the regime of small $m_a$, or 
\be
m_a\lesssim\frac{3^{{7}/{5}}}{2^{{4}/{5}}5^{{1}/{2}}7^{{1}/{10}}}\left(\frac{m_e^2\omega^2}{\alpha R^3B_0|\sin\theta|}\right)^{{1}/{5}},
\ee
which becomes
\be
\label{maboun}
m_a\lesssim\frac{3^{{7}/{5}}\Gamma\!\left(\frac{2}{5}\right)^2}{2^15^{{1}/{2}}7^{{1}/{10}}\pi\Gamma\!\left(\frac{4}{5}\right)}\left(\frac{m_e^2\omega^2}{\alpha R^3B_0}\right)^{{1}/{5}}\approx1.45\times10^{-5}\,\text{eV}\left(\frac{\omega}{1\,\text{keV}}\right)^{{2}/{5}}\left(\frac{R}{10\,\text{km}}\right)^{-{3}/{5}}\left(\frac{B_0}{B_c}\right)^{-{1}/{5}},
\ee
once averaged over the angle $\theta$ as is already done for $p_a$. 

The results in this subsection are analytical approximations, valid in the appropriate regime, which are valuable for more clearly displaying how the different physical processes affect the final Stokes parameters. In the rest of this paper, all results are obtained from the explicit numerical computations of ALP luminosity, as detailed in Section \ref{SProd} (involving the proper integration over the magnetar profiles to generate maximum and minimum luminosity limits based on different EoSs, magnetar masses, and superconductivity assumptions) and explicit numerical computations of the evolution equations [cf.~Eqs.~\eqref{EqnEE}], which allow full treatment of the ALP-inclusive Stokes parameters up to high ALP mass, without needing the approximations made in this subsection.

%%%%%%%%%%%%%%%%%%%%%%%%%%%%%%%%%%%%%%%%%%%%%%%%%%%%%%%%%%%%%%%%%

\section{{\it IXPE} data and experimental considerations}
\label{sec:data}

\begin{table*}
	\centering
    \small
     \tabcolsep=0.11cm
	\begin{tabular}{c c c c c c  } % four columns, alignment for each
		\hline
                Energy bin (keV) & $I$ (counts) & $Q$ (counts) & $U$ (counts) & PD (\%) & PA (deg) \\
            \hline
                2.00\,-\,3.00 & $19960$ & $-175.6$ & $3161$ & $15.86$ & $46.59$  \\
                3.00\,-\,4.00 & $6112$  & $-231.5$ & $847.2$ & $14.37$ & $52.64$  \\
                4.00\,-\,4.80 & $1630.$ & $12.59$ & $96.16$ & $5.949$ & $41.27$  \\
                4.80\,-\,5.50 & $641.1$ & $-38.80$ & $-48.36$ & $9.669$ & $-64.37$ \\
                5.50\,-\,8.00 & $687.7$ & $47.92$  & $-237.2$ & $35.19$ & $-39.29$ \\
		\hline
	\end{tabular}
    	\caption{
            Condensed version of the observational {\it IXPE} data for 4U 0142+61, with the full version of the data files available on NASA's HEASARC \cite{HEASARC}. The values for Stokes $I$ were obtained from our re-processing of the data file. Then, PD and PA were obtained by extracting the experimentally-obtained central values (the crosses) in Figure 2 of Ref.~\cite{2022Sci...378..646T}. Finally, $Q$ and $U$, given with respect to celestial north, are derived from the extracted PD and PA values according to Eqs.~\eqref{EqnPD} and \eqref{EqnPA}. 
    	}   
        \label{tab:mag1exp}
\end{table*}

\begin{table*}
	\centering
    \small
     \tabcolsep=0.11cm
	\begin{tabular}{c c c c c c  } % four columns, alignment for each
		\hline
                Energy bin (keV) & $I$ (counts) & $Q$ (counts) & $U$ (counts) & PD (\%) & PA (deg) \\
            \hline
                2.00\,-\,3.00 & $5876$  & $-712.9$ & $-1028$  & $21.29$ & $-62.37$ \\
                3.00\,-\,4.00 & $2527$  & $-593.2$ & $-852.8$ & $41.11$ & $-62.41$ \\
                4.00\,-\,5.00 & $1103$  & $-356.0$ & $-537.0$ & $58.41$ & $-61.77$ \\
                5.00\,-\,6.00 & $550.8$ & $-164.8$ & $-271.4$ & $57.65$ & $-60.63$ \\
                6.00\,-\,8.00 & $514.5$ & $-230.6$ & $-374.5$ & $85.48$ & $-60.81$ \\
		\hline
	\end{tabular}
    	\caption{
            Condensed version of the observational {\it IXPE} data for 1RXS J170849.0-400910, with the full version of the data files available on NASA's HEASARC data archive \cite{HEASARC}. The values for Stokes $I$ were obtained from our re-processing of the data file. Then, PD and PA were obtained by extracting the experimentally-obtained central values (the dots) from the right panel of Figure 1 in Ref.~\cite{2023ApJ...944L..27Z}. Finally, $Q$ and $U$ are derived from the extracted PD and PA values according to Eqs.~\eqref{EqnPD} and \eqref{EqnPA}. 
    	}   
        \label{tab:mag2exp}
\end{table*}

The polarization properties of a beam of light can be completely described by the four Stokes parameters: $I$ (total intensity), $Q$ and $U$ (linear polarizations), and $V$ (circular polarization)~\cite{1954AmJPh..22..351M}. In the detector frame, Stokes $Q_\text{obs}^\text{det}$ gives the amount of linear polarization in the celestial north-south ($+Q$) or east-west ($-Q$) directions, and Stokes $U_\text{obs}^\text{det}$ gives the amount of linear polarization in the direction 45$^{\circ}$ east-of-celestial north ($+U$) or 45$^{\circ}$ west-of-celestial north ($-U$) (using the convention of Ref.~\cite{2022Sci...378..646T}).\footnote{ 
In the rest of this section, as well as in the tables, we drop the superscript and subscript in $Q_\text{obs}^\text{det}$ and $U_\text{obs}^\text{det}$, since the context is clear.}  
$Q$ and $U$ can then be combined according to the prescriptions of Ref.~\cite{2015APh....68...45K} to yield the (linear) polarization degree (hereafter, just `PD') and polarization angle (`PA'):
\be\label{EqnPD}
\text{PD} = \frac{\sqrt{Q^2+U^2}}{I}\,\,,
\ee
\be\label{EqnPA}
\text{PA} = \frac{1}{2}\,\arctan\left(\frac{U}{Q} \right)\,\,.
\ee
Since {\it IXPE} is only able to detect linear $X$-ray polarization, its data contain the number of Stokes $I$, $Q$, and $U$ counts at energies throughout its 2~--~8~keV energy bandpass. Tables~\ref{tab:mag1exp} and \ref{tab:mag2exp} summarize these experimental/observational data for 4U 0142+61 and 1RXS J170849.0-400910, respectively.

The {\it IXPE} discovery papers for these two magnetars~\cite{2022Sci...378..646T, 2023ApJ...944L..27Z} provide astrophysical modelling, based on various models of thermal emission from the surface, as well as the standard resonant cyclotron scattering (RCS) of that emission afterwards in the magnetosphere, to explain the general trends of polarization degree and angle reported by the instrument. Readers interested in further details of these mechanisms, as well as the specific results from this modelling, are referred to those papers. For the purposes of our study, we assume that the astrophysical modelling will eventually be able to completely describe the observational results and align identically with the experimental results found for (PD, PA). Then, since the ALP-to-photon conversions occur at a distance (on the order of 100 times the radius $R$ of the magnetar, see Ref.~\cite{Fortin:2018aom}) far beyond the radius of the final astrophysical scattering (about 10 times the radius $R$, see Ref.~\cite{2020MNRAS.492.5057T}), any possible ALP contribution will simply modify the final astrophysical (PD, PA). 

\begin{figure}
  \centering
\includegraphics[width=0.7\textwidth]{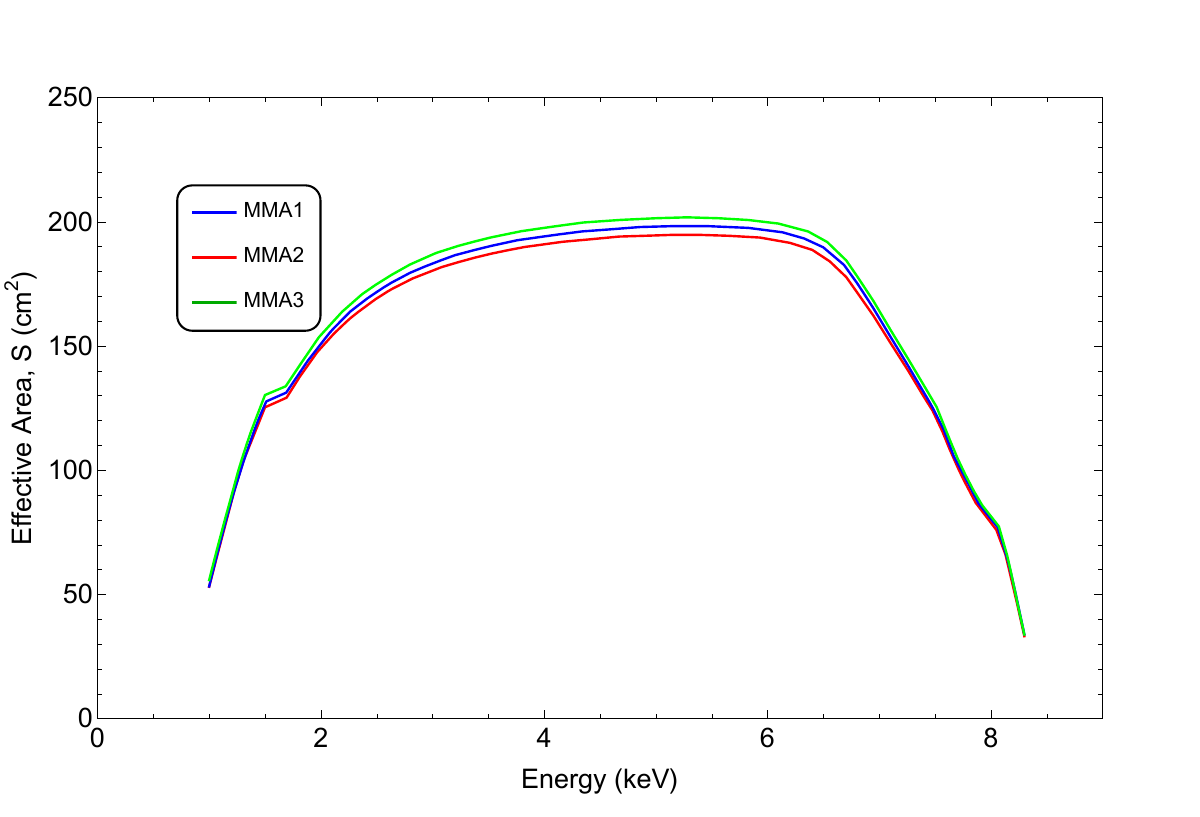}
\caption{The best-fit models of the effective area of each of the three Mirror Module Assemblies (MMAs, {\it i.e.}, the three detectors) of {\it IXPE} as a function of photon energy. This plot is re-made from Figure 8 in Ref.~\cite{2022JATIS...8b6002W}. }
  \label{fig:detector}
\end{figure}

Among the various detector responses involved, the effective area (which includes the detector efficiency) is the main one that this work must account for.\footnote{Other detector effects include the energy redistribution of the instrument, as well as the energy dependence of the modulation factor. However, the effect of the former is not significant for us, given that the width of the energy bins we use is large in comparison to the amount of redistribution~\cite{2022JATIS...8b6002W}. The effect of the latter was already accounted for in the polarization results reported by {\it IXPE} (and, indeed, must already be adjusted for in any reported polarization result). For more details on how to recover the polarization degree using the modulation amplitude in an experimental context, see  Ref.~\cite{2014JAI.....340008B}.} Thus, Fig.~\ref{fig:detector} displays the effective areas of {\it IXPE}'s three detectors~\cite{2022JATIS...8b6002W}. Our full numerical calculations incorporate this energy-dependent effective area as the area of the detector, in a manner similar to the inclusion of $S$ in Eq.~(\ref{Eqpa}). 

The {\it IXPE} observational data for the (PD, PA) values are summarized in Fig.~\ref{fig:exp1} for the magnetar 4U 0142+61 and in Fig.~\ref{fig:exp2} for the magnetar RXS J170849.0-400910. The data are grouped into energy bins between 2~--~8~keV and displayed on a polar plot, with PD as the radial coordinate and PA as the azimuthal coordinate.
% to ensure sufficient counts in each bin. 
The central data point for each bin is shown with a red dot, and the enclosing blue contours depict the 68.3\% confidence level (C.L.) regions in Fig.~\ref{fig:exp1} obtained for the case of 4U 0142+61 and the 50\%-C.L. regions in Fig.~\ref{fig:exp2} obtained for the case of RXS J170849.0-400910. We refer the respective discovery papers to readers interested in further experimental details~\cite{2022Sci...378..646T,2023ApJ...944L..27Z}.

%%%%%%%%%%%%%%%%%%%%%%%%%%%%%%%%%%%%%%%%%%%%%%%%%%%%%%%%%%%%%%%%%

\section{Constraint procedure and results}
\label{sec:results}

\begin{figure}
  \centering
\includegraphics[width=0.6\textwidth]{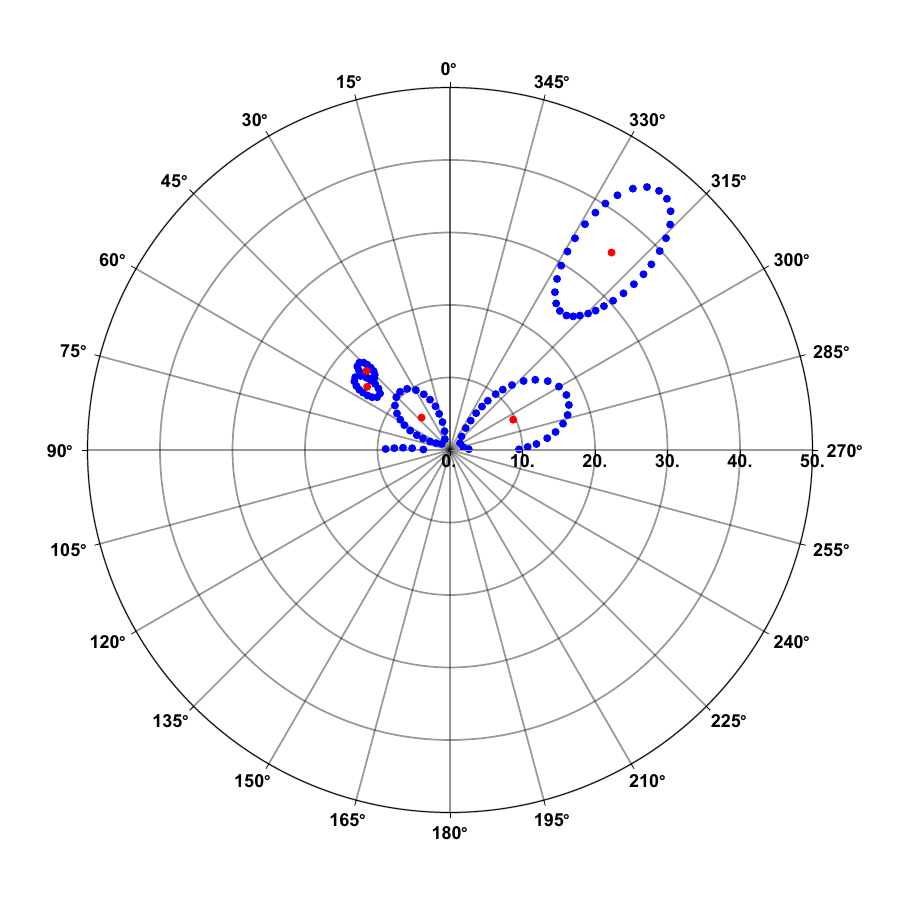}
\caption{Polar plot for the magnetar 4U 0142+61, which displays the experimental central values (red) and 68\% C.L. uncertainty bounds (blue) for those values. The radial coordinate denotes the PD in percentage; the angular coordinate denotes the PA in degrees. We note that $0^{\circ}$, or celestial north, is to the top of the figure. We also note that that, due to the $180^{\circ}$ oscillation of the electromagnetic field, the lower half of the circular plane (or angles past $90^{\circ}$ on either side of celestial north) is degenerate with the top half, thus causing seeming discontinuities in one or two of the energy bins. These values have been digitized from Figure 2 of Ref.~\cite{2022Sci...378..646T}.}
  \label{fig:exp1}
\end{figure}

\begin{figure}
  \centering
\includegraphics[width=0.6\textwidth]{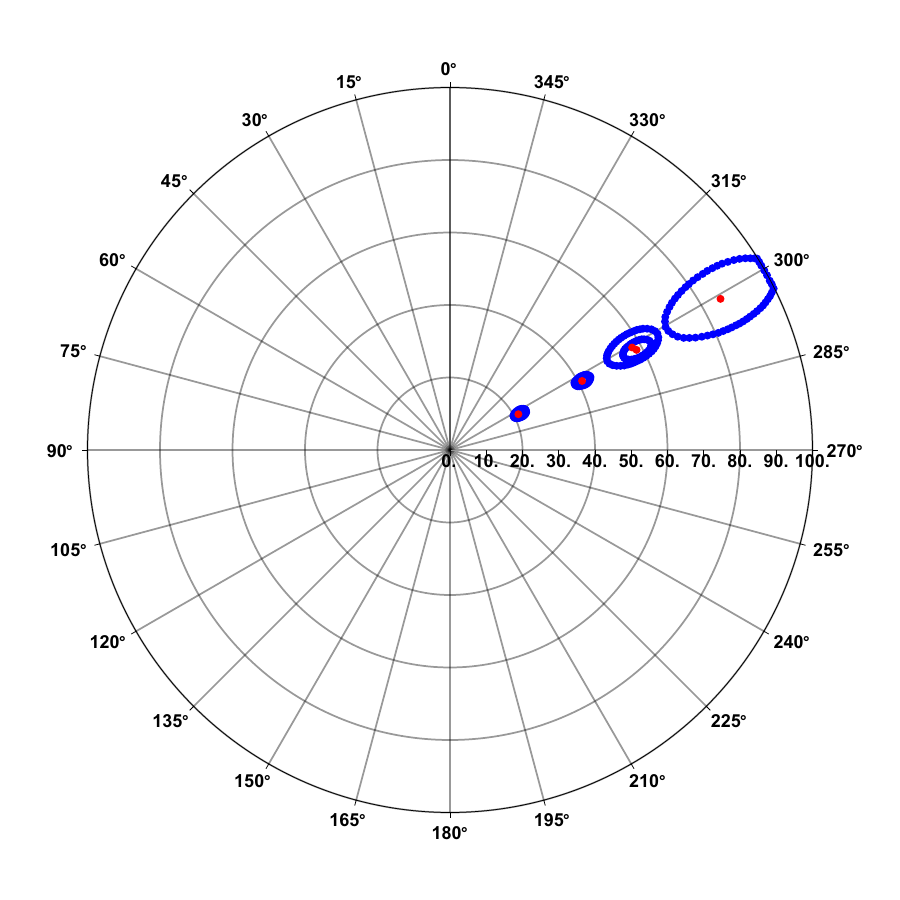}
\caption{Polar plot for the magnetar 1RXS J170849.0-400910, which displays the experimental central values (red) and 50\% C.L. uncertainty bounds (blue) for those values. The radial coordinate denotes the PD in percentage; the angular coordinate denotes the PA in degrees. We note that $0^{\circ}$, or celestial north, is to the top of the figure. These values have been digitized from Figure 1 of Ref.~\cite{2023ApJ...944L..27Z}.}
  \label{fig:exp2}
\end{figure}

\begin{figure}
  \centering
\includegraphics[width=0.4\textwidth]{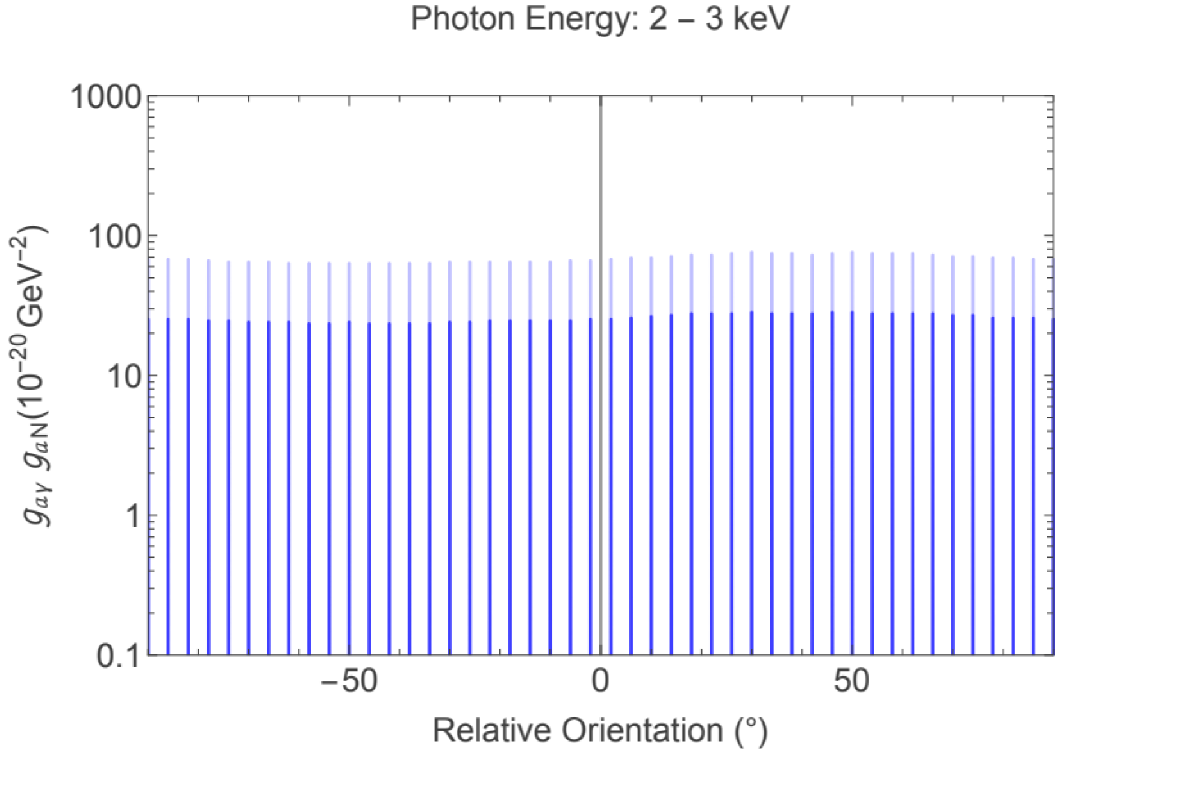}\includegraphics[width=0.4\textwidth]{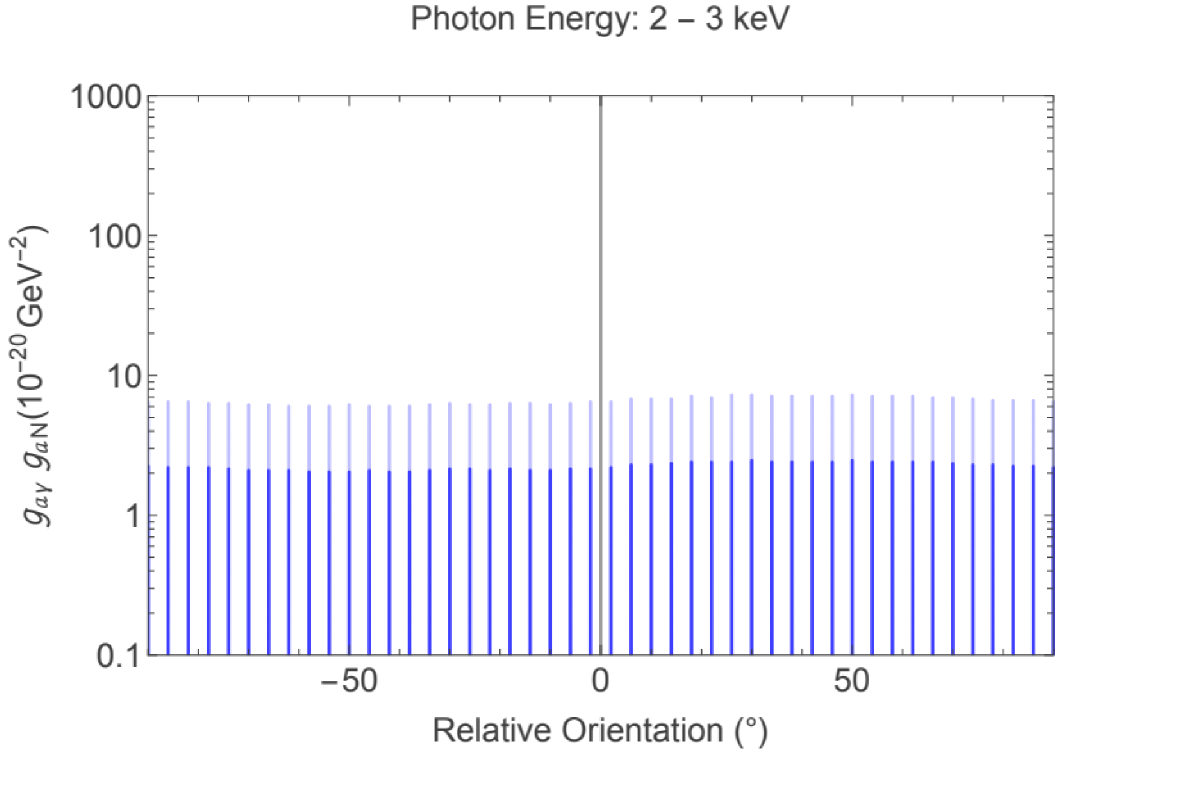}\\
\includegraphics[width=0.4\textwidth]{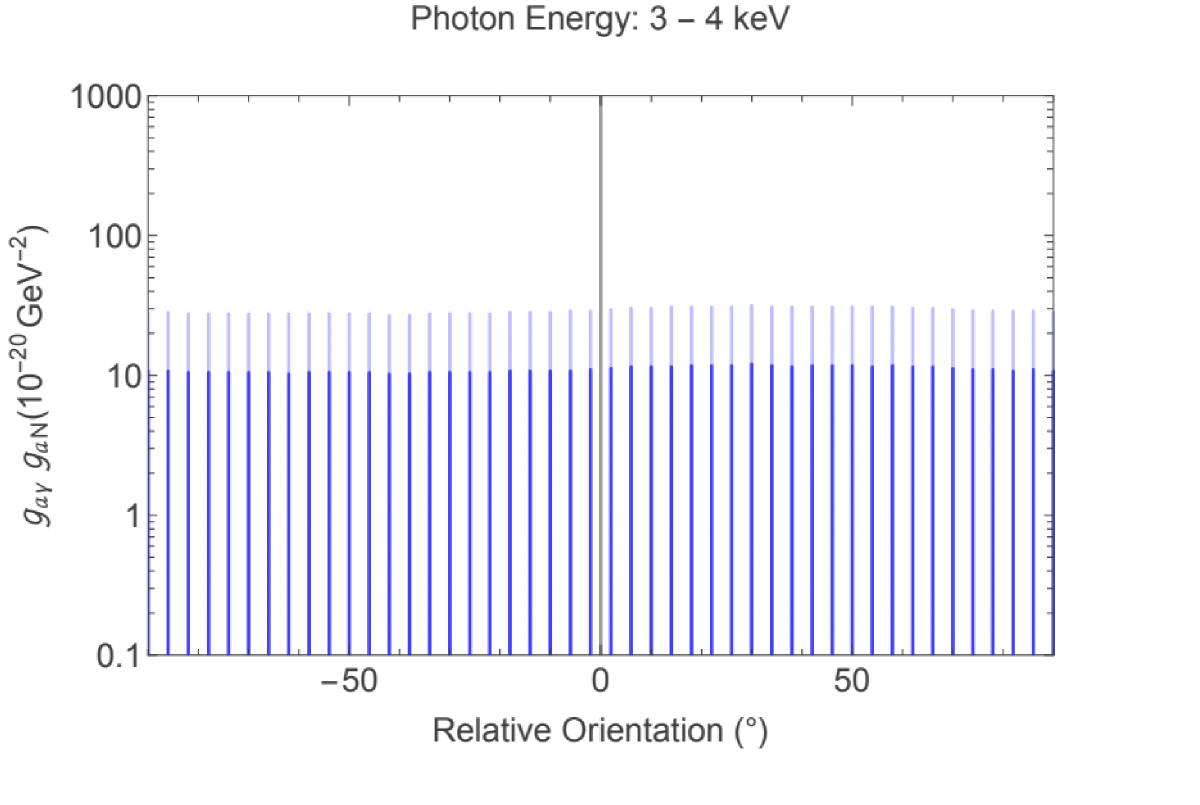}\includegraphics[width=0.4\textwidth]{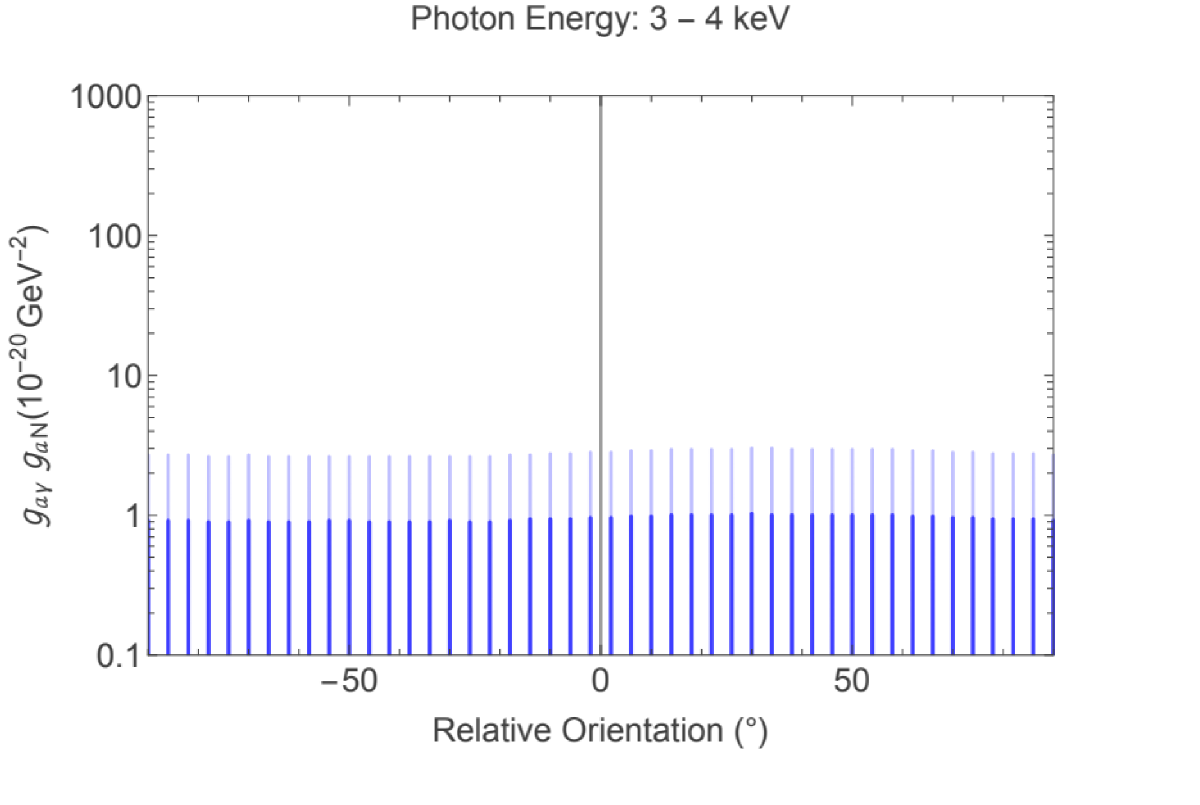}\\
\includegraphics[width=0.4\textwidth]{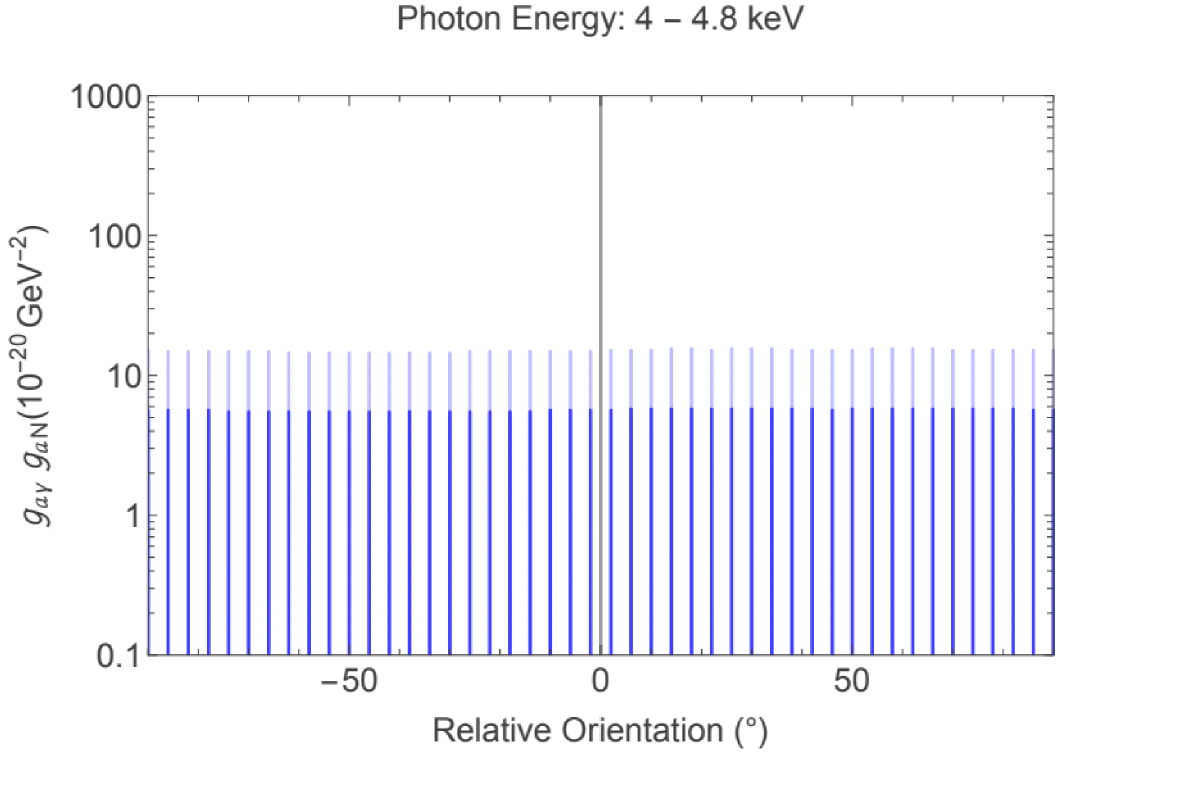}\includegraphics[width=0.4\textwidth]{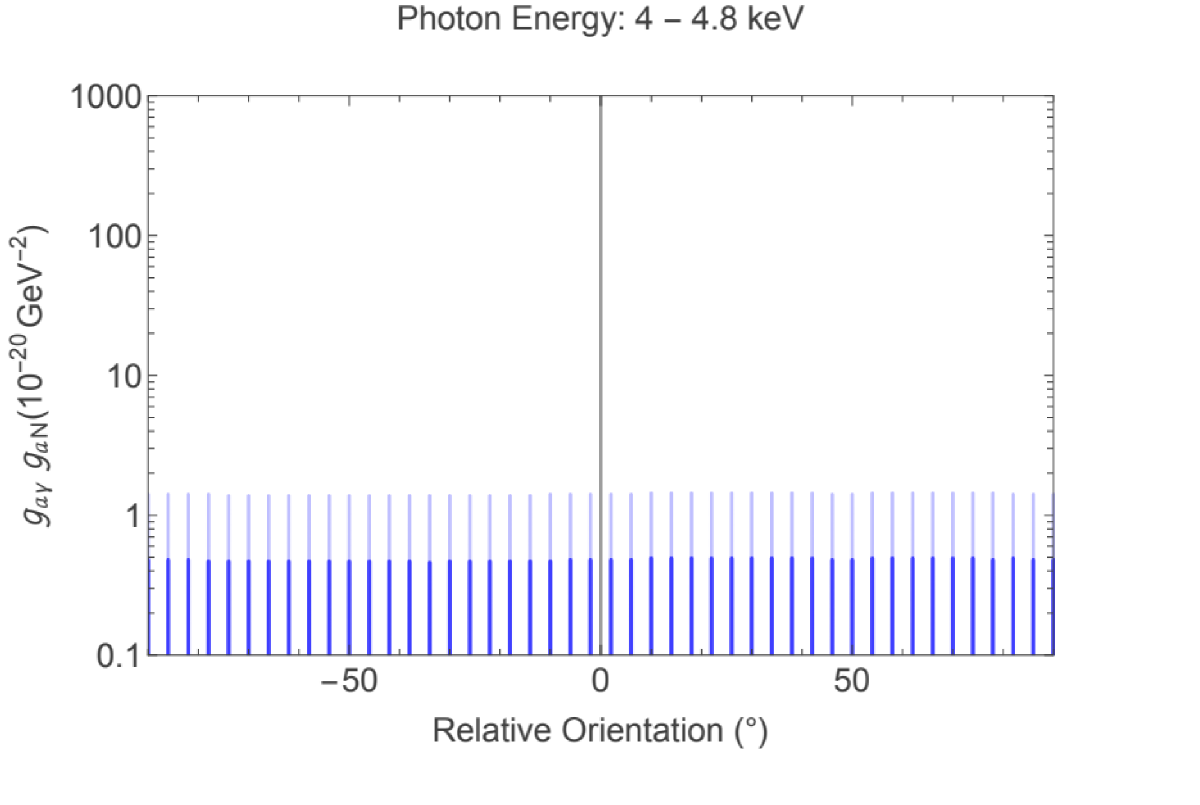}\\
\includegraphics[width=0.4\textwidth]{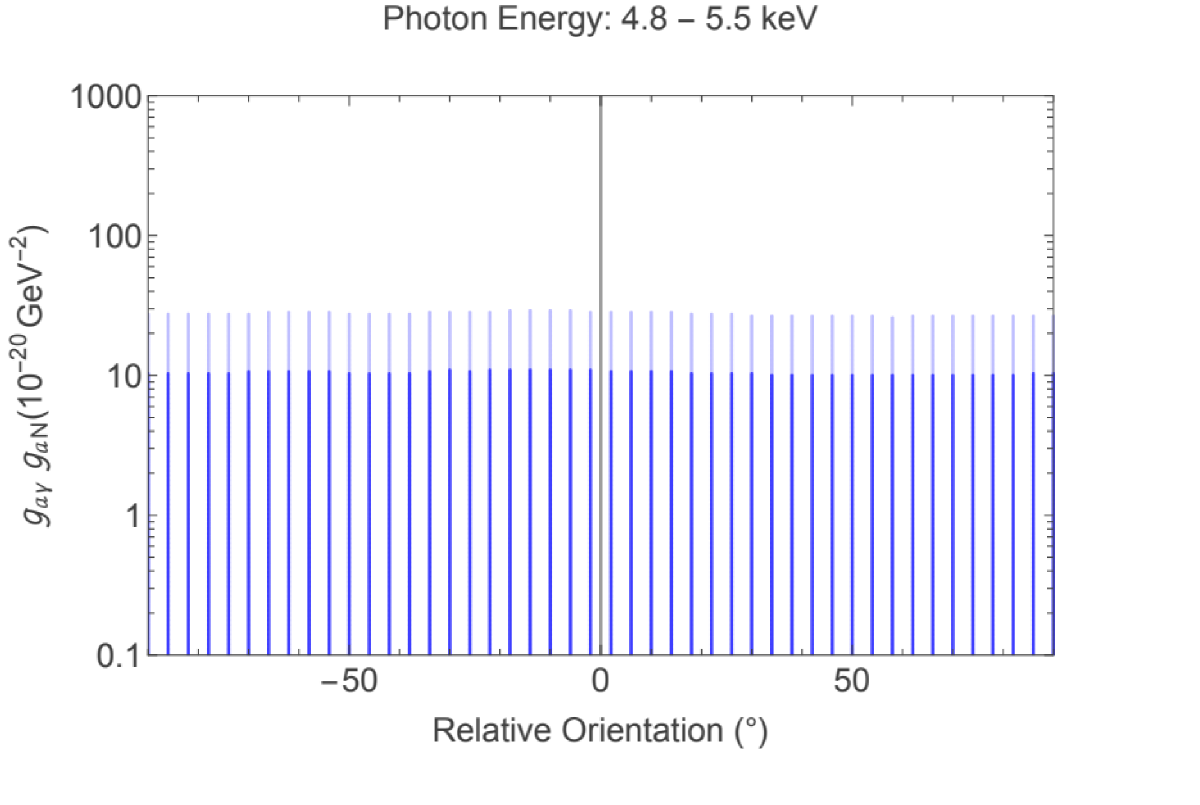}\includegraphics[width=0.4\textwidth]{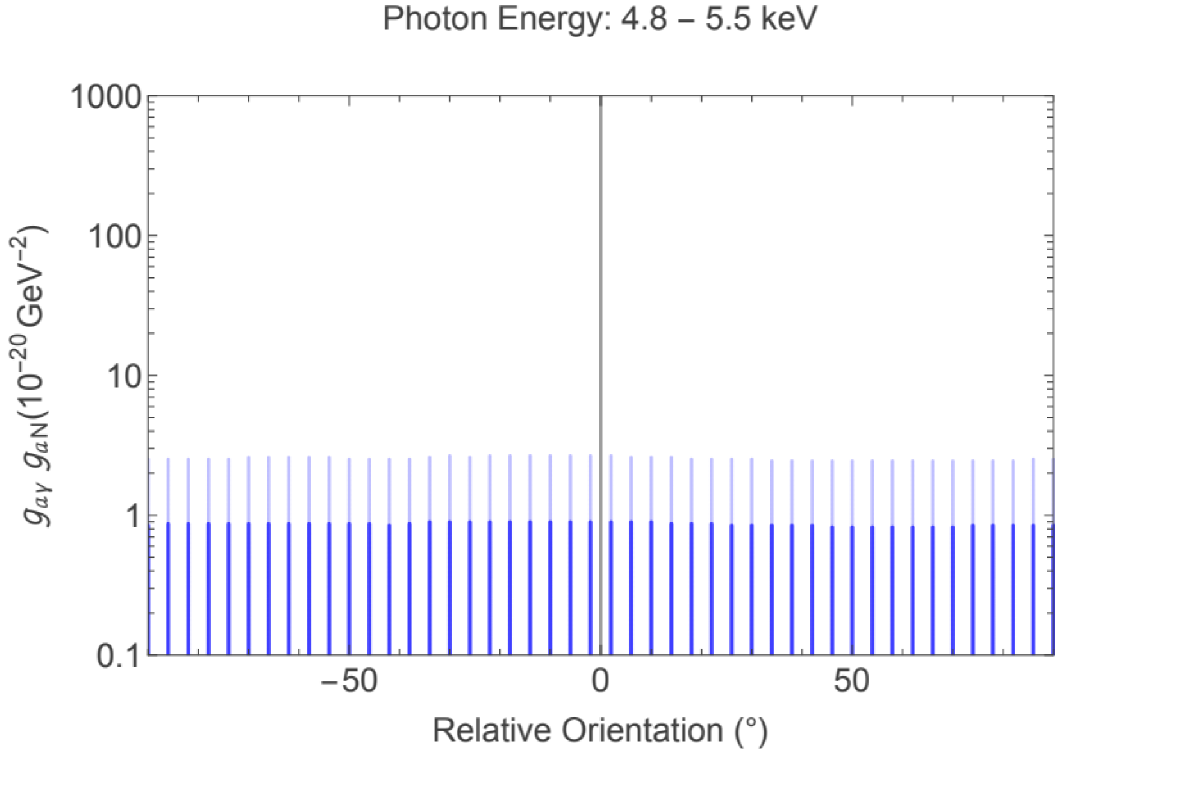}\\
\includegraphics[width=0.4\textwidth]{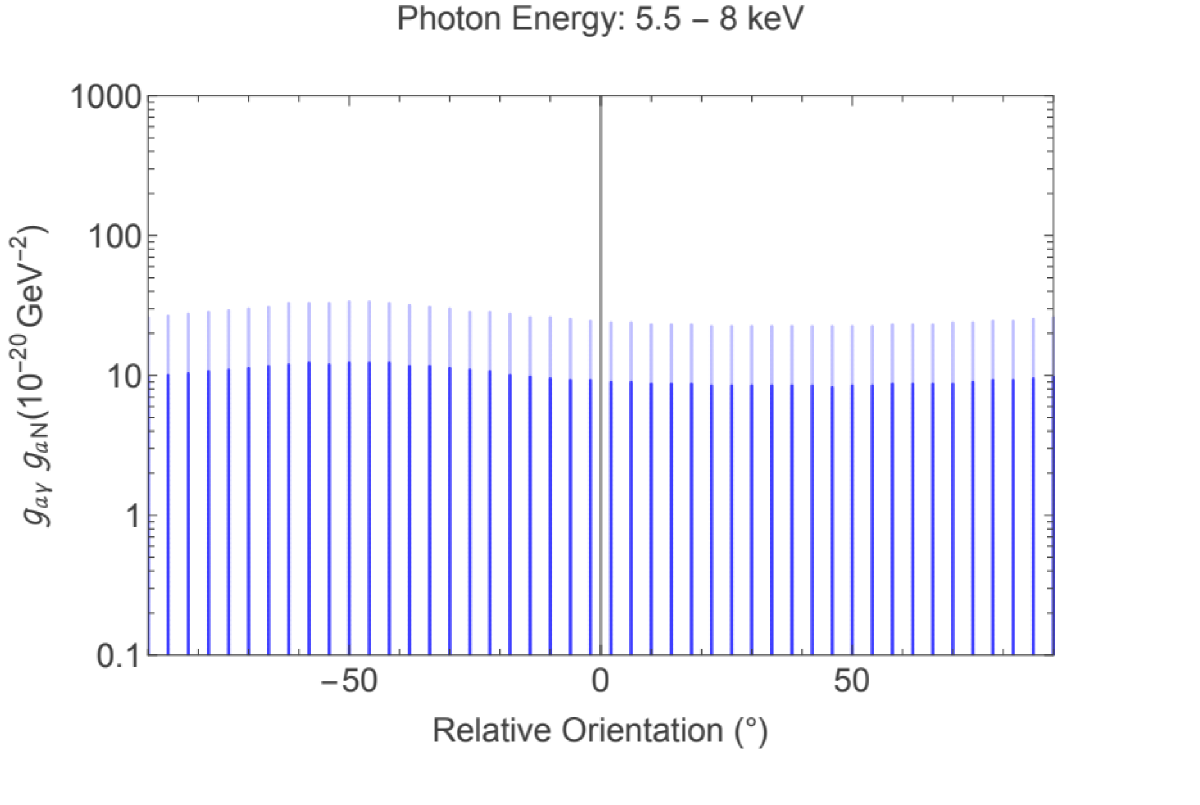}\includegraphics[width=0.4\textwidth]{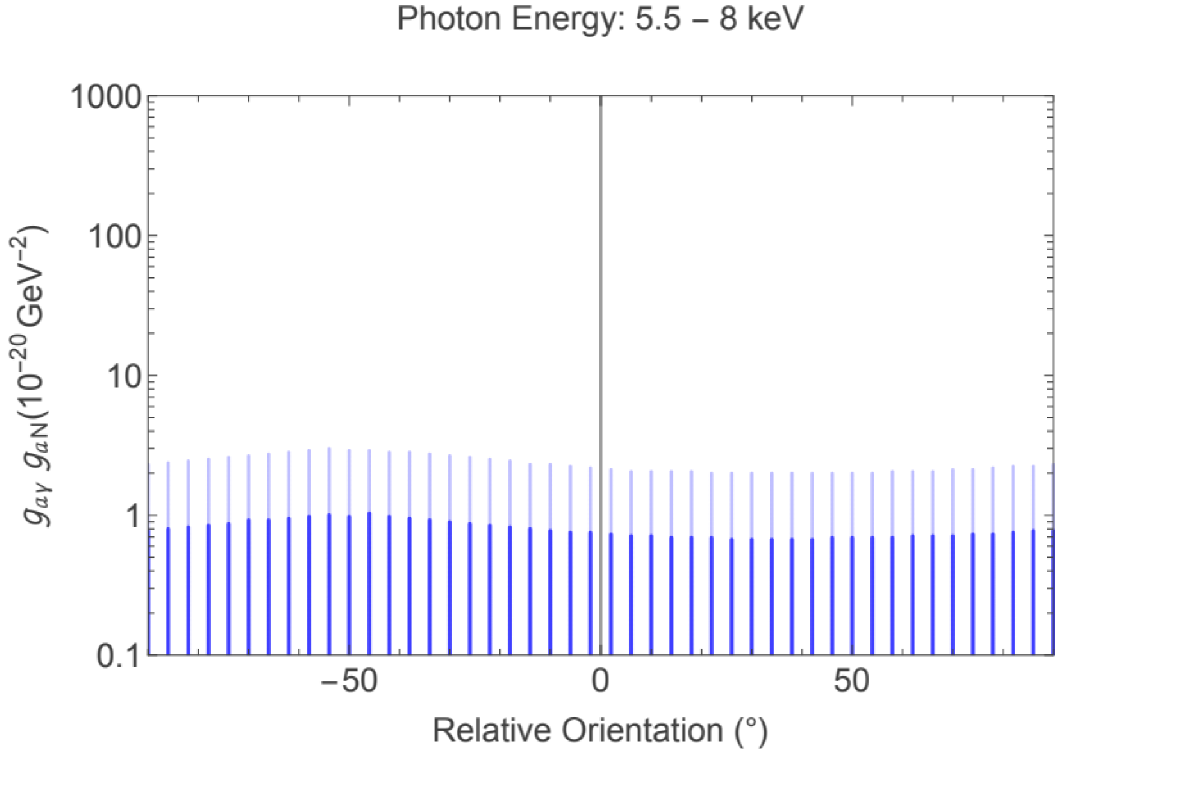}
  \caption{Plots for allowed values of $g_{a\gamma}g_{aN}$ for all possible relative orientations $\Delta \psi$ of the magnetar 4U 0142+61. Results using a core temperature of $T^\infty = 1~(5) \times 10^8$~K are shown in the left (right) column. Results corresponding to the most optimistic and conservative luminosities (obtained by varying the EoSs and magnetar masses) for a given core temperature are displayed with brighter and fainter colors, respectively, in each panel. The different panels correspond to different energy bins.}
  \label{fig:constblob1}
\end{figure}

\begin{figure}
\centering
\includegraphics[width=0.4\textwidth]{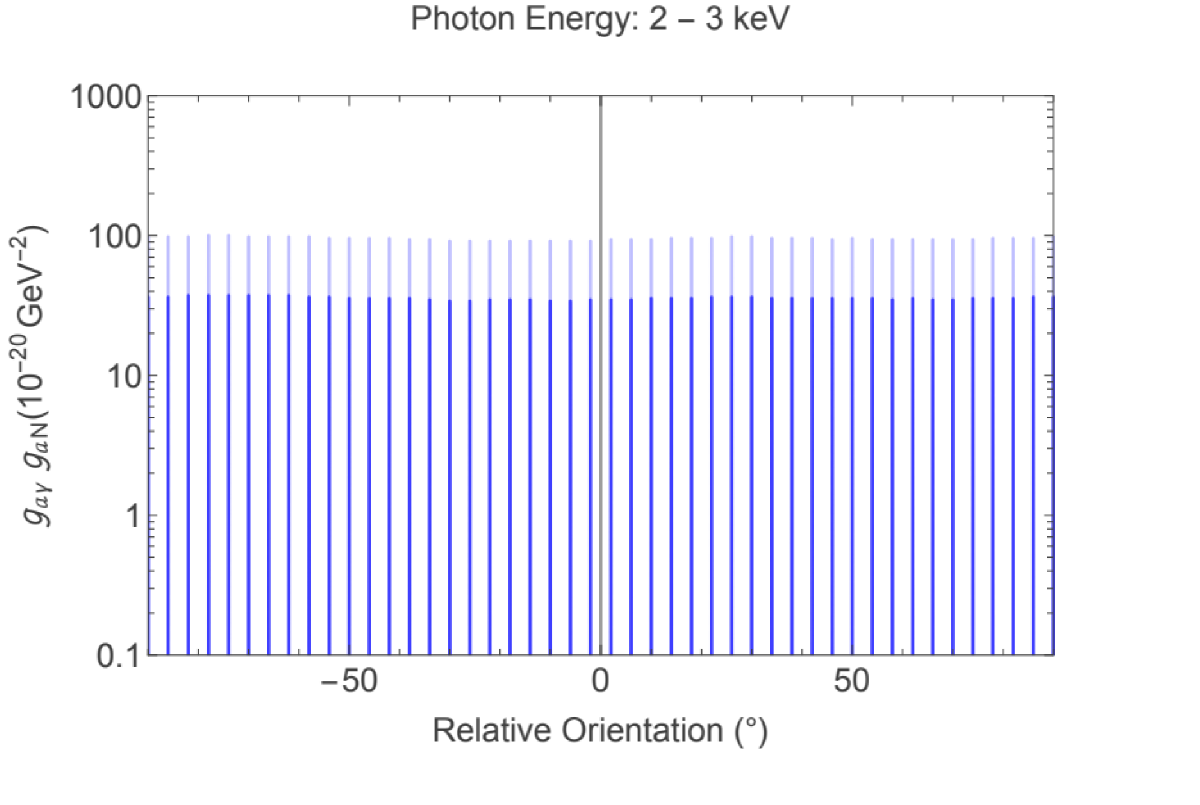}\includegraphics[width=0.4\textwidth]{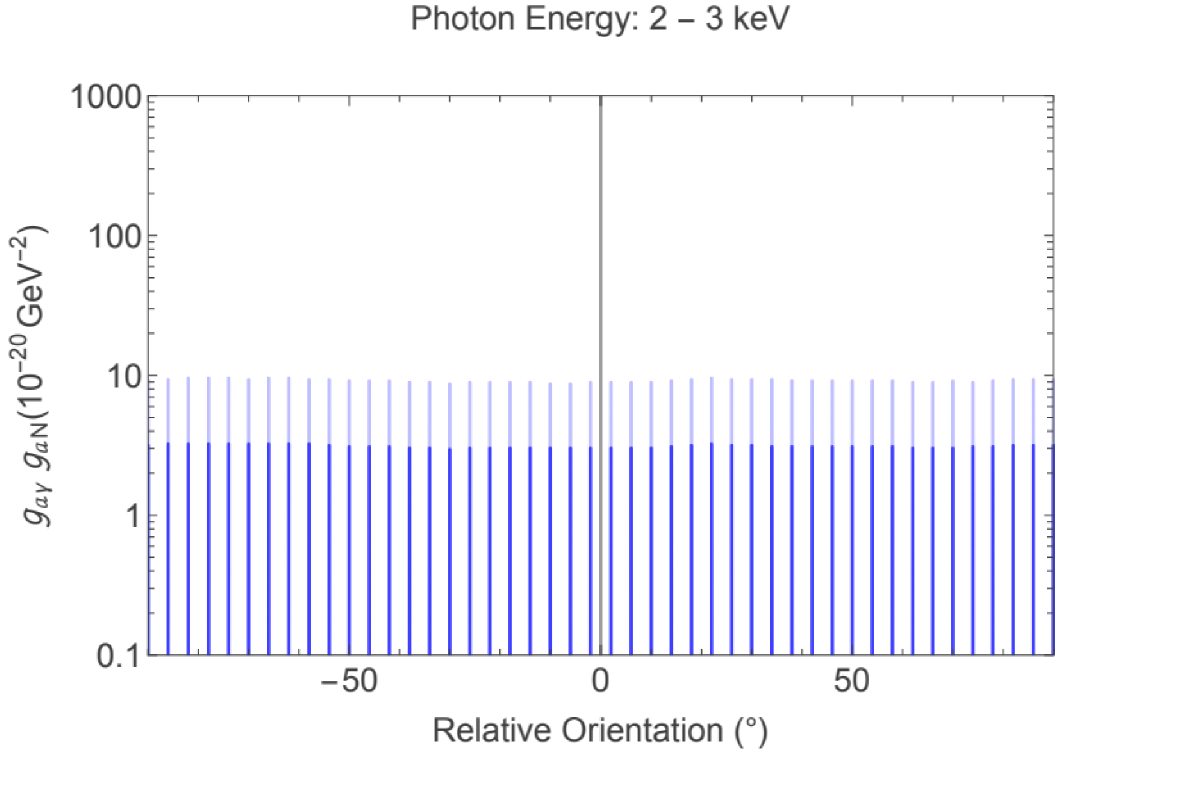}\\
\includegraphics[width=0.4\textwidth]{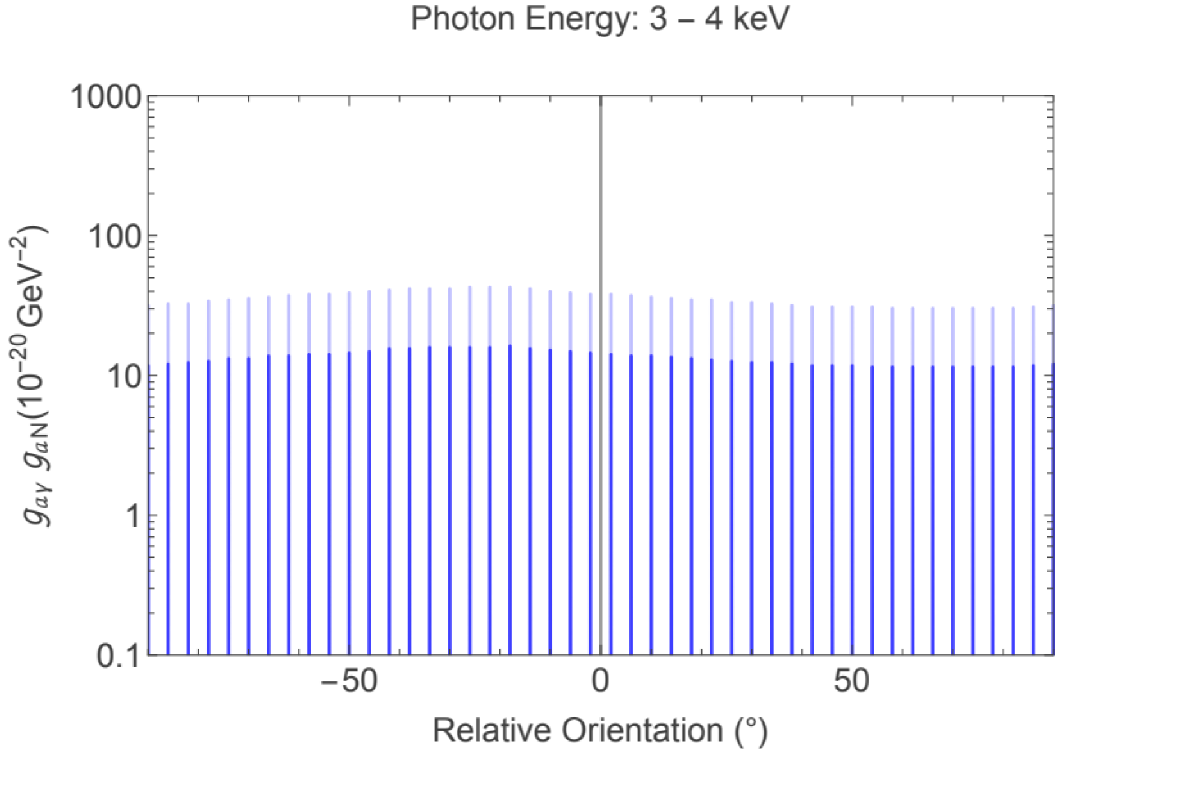}\includegraphics[width=0.4\textwidth]{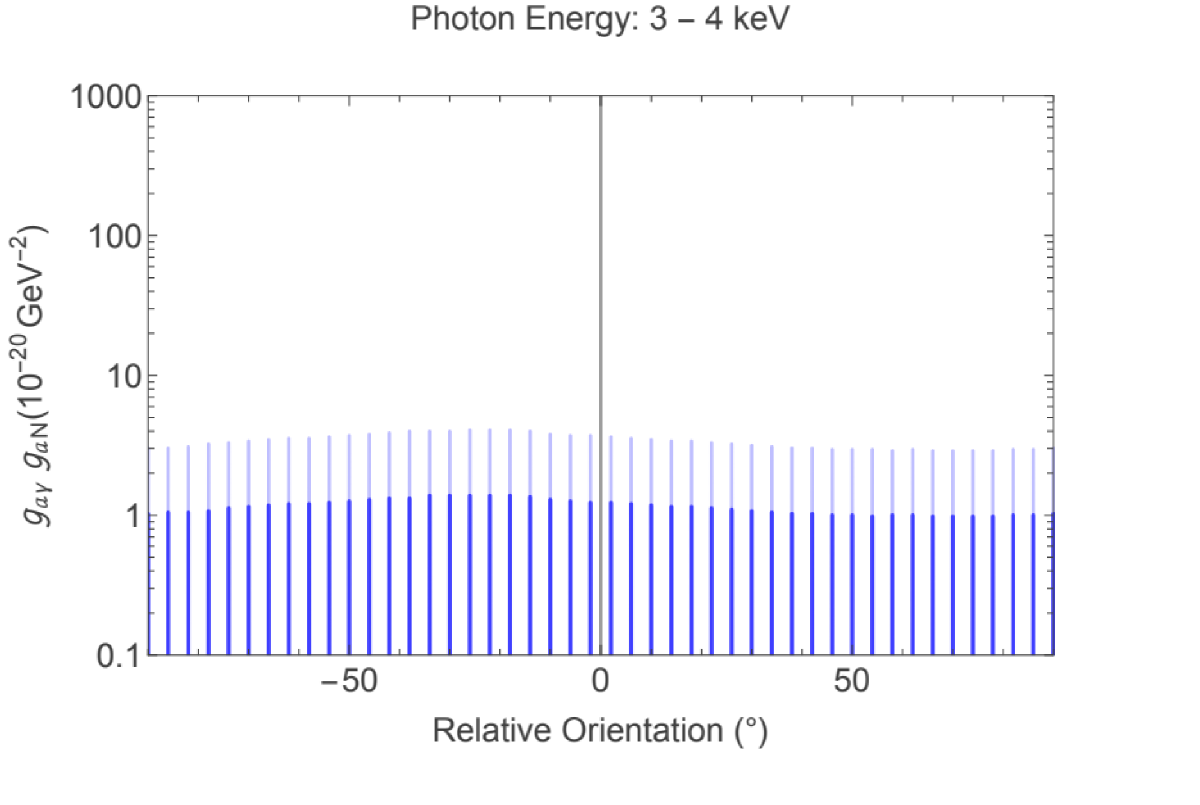}\\
\includegraphics[width=0.4\textwidth]{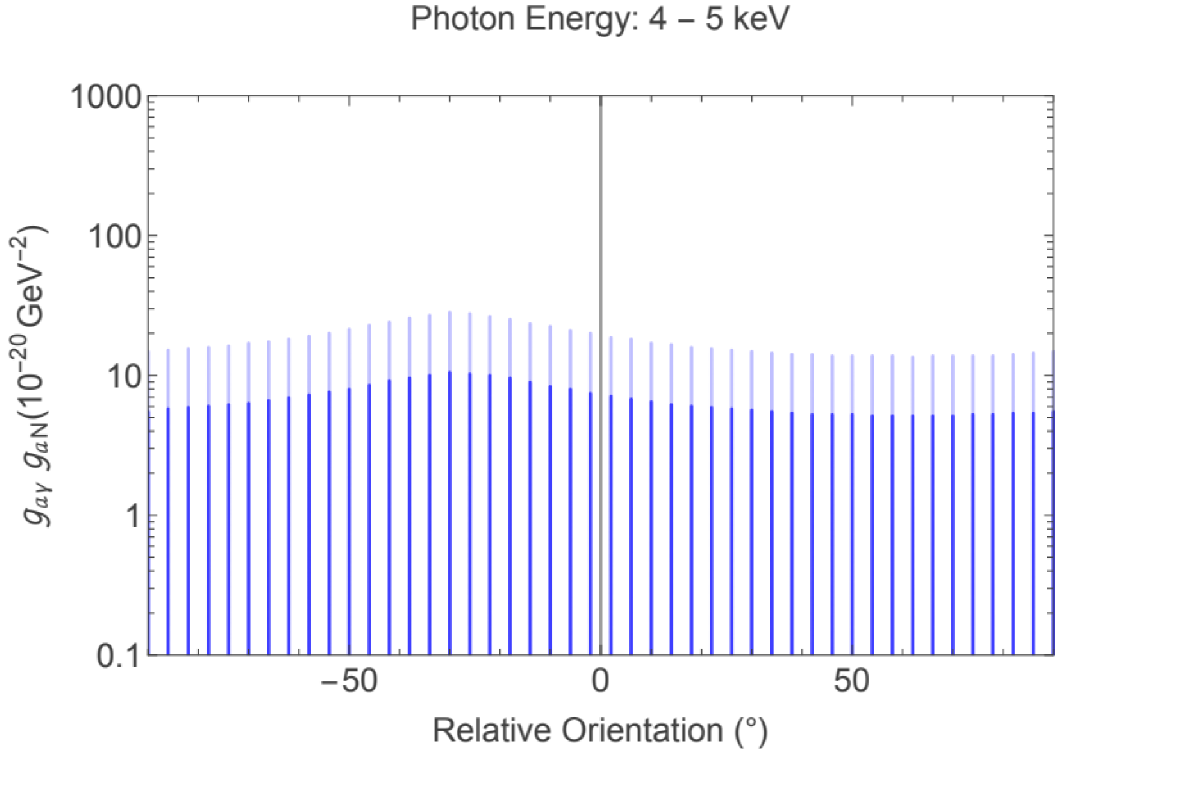}\includegraphics[width=0.4\textwidth]{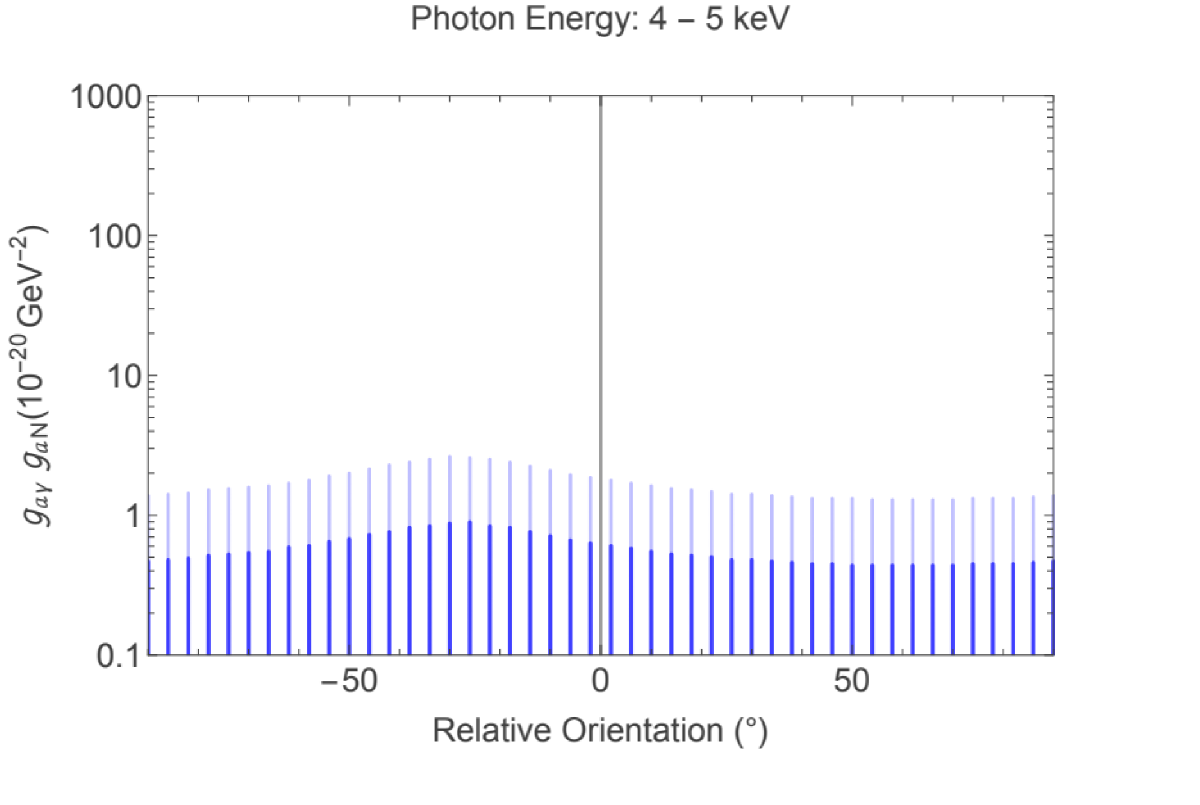}\\
\includegraphics[width=0.4\textwidth]{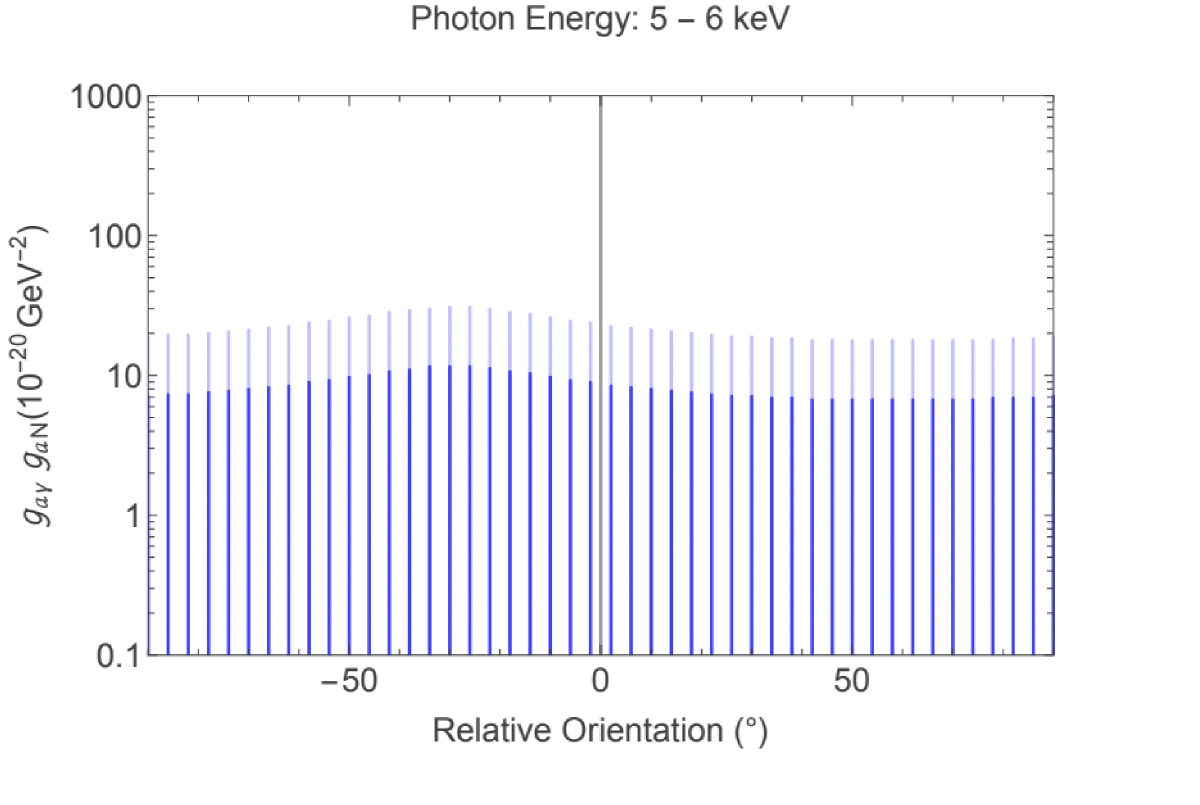}\includegraphics[width=0.4\textwidth]{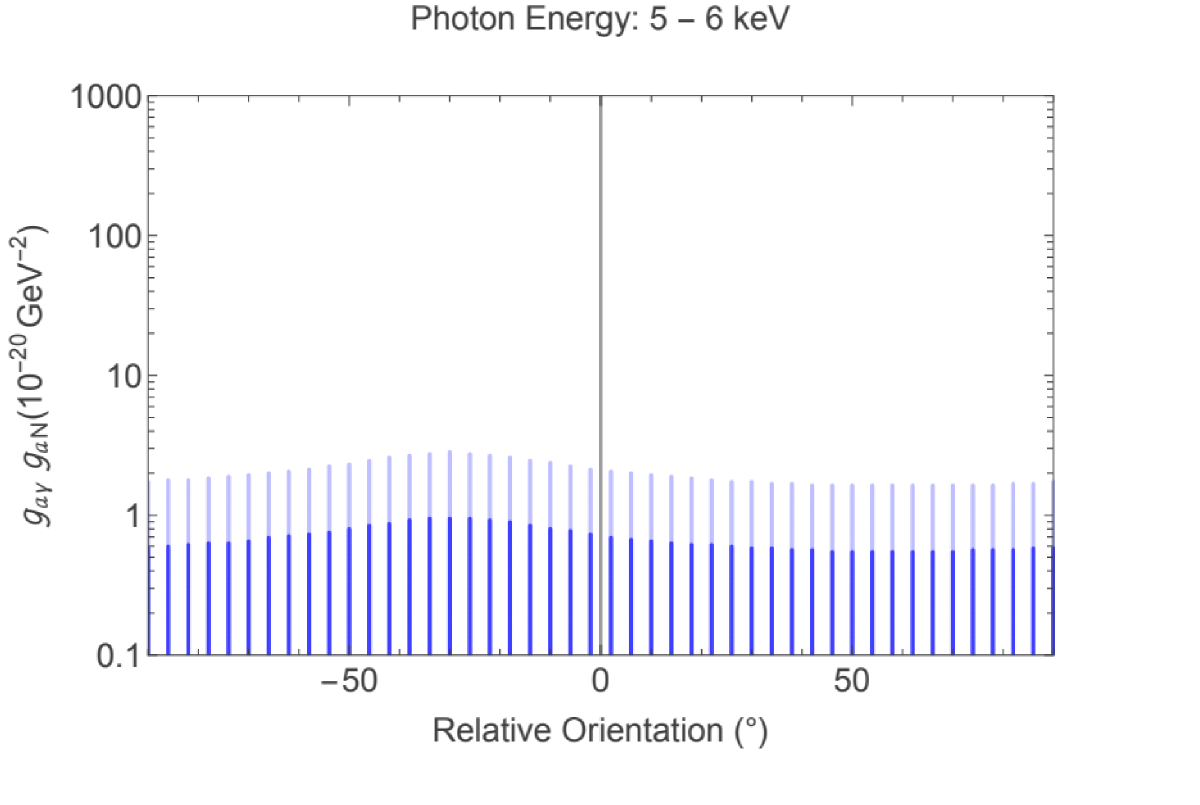}\\
\includegraphics[width=0.4\textwidth]{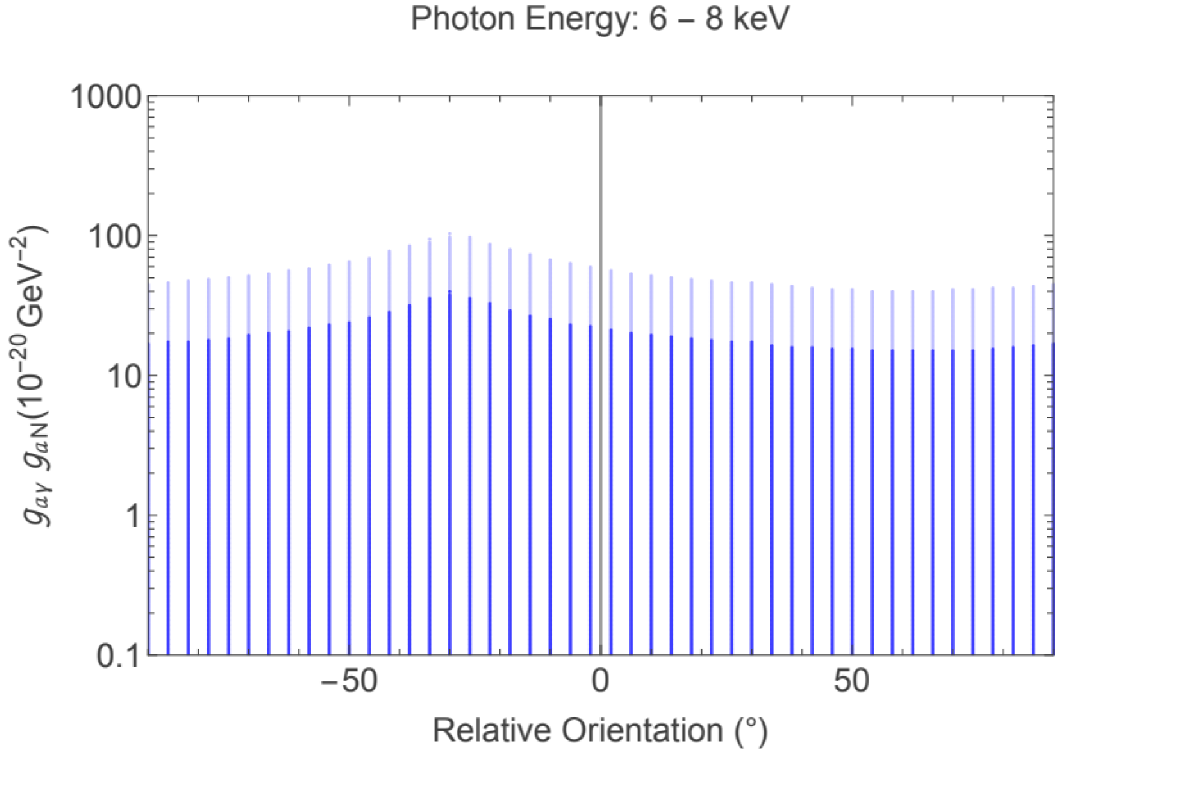}\includegraphics[width=0.4\textwidth]{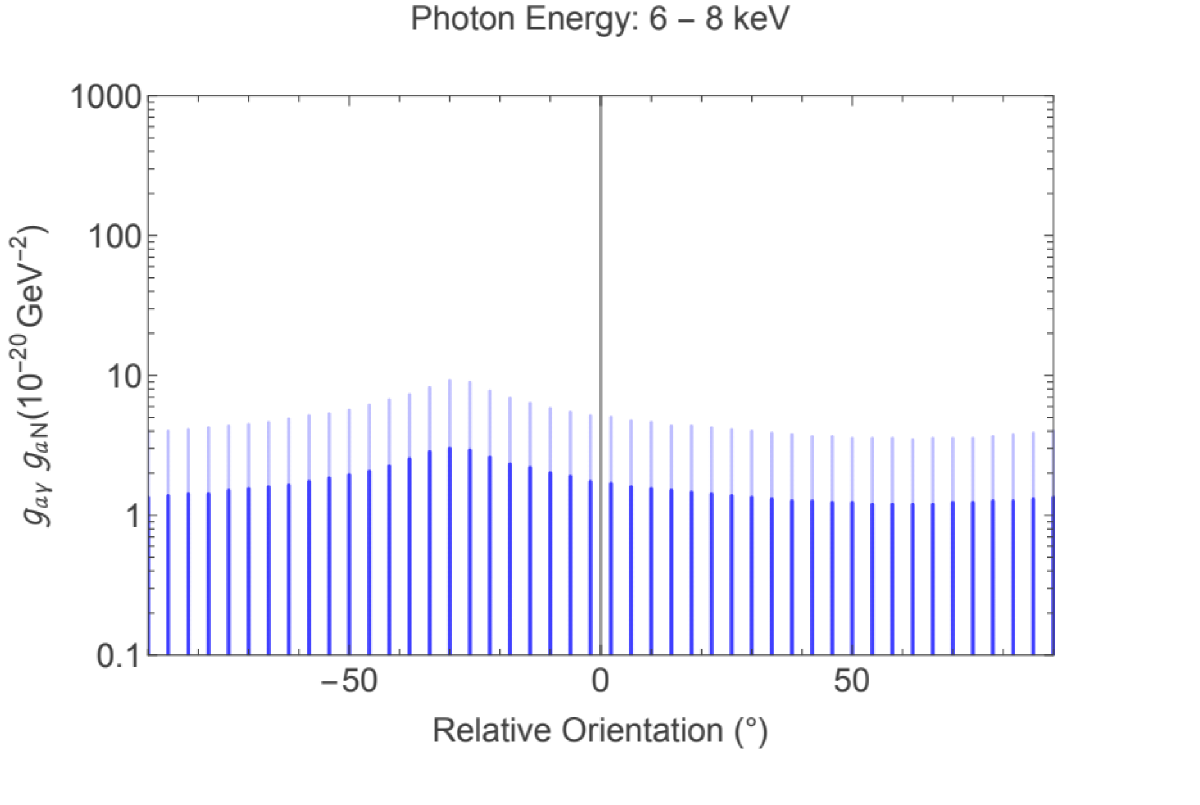}
    \caption{Plots for allowed values of $g_{a\gamma}g_{aN}$ for all possible relative orientations $\Delta \psi$ of the magnetar 1RXS J170849.0-400910. Results using a core temperature of $T^\infty = 1~(5) \times 10^8$~K are shown in the left (right) column. Results corresponding to the most optimistic and conservative luminosities (obtained by varying the EoSs and magnetar masses) for a given core temperature are displayed with brighter and fainter colors, respectively, in each panel. The different panels correspond to different energy bins.}
    \label{fig:constblob2}
\end{figure}

\begin{figure}
  \centering
\includegraphics[width=0.49\textwidth]{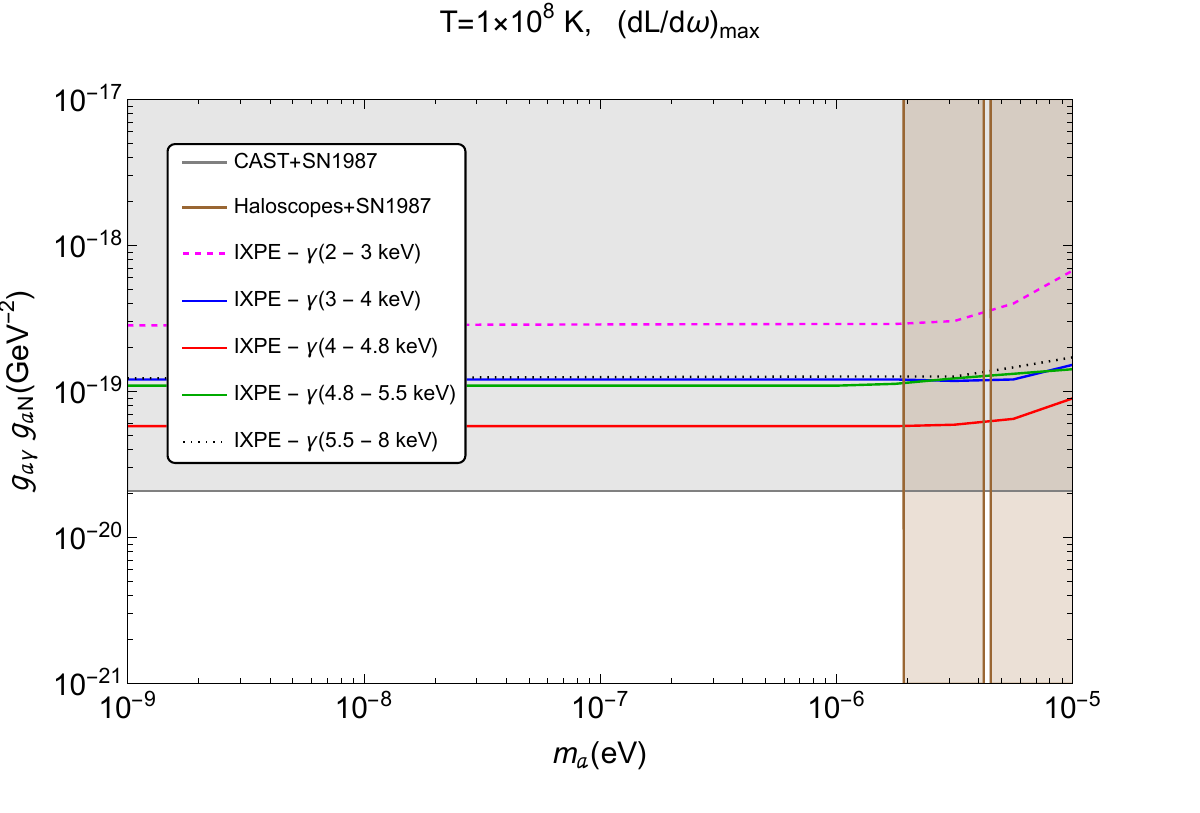}
\includegraphics[width=0.49\textwidth]{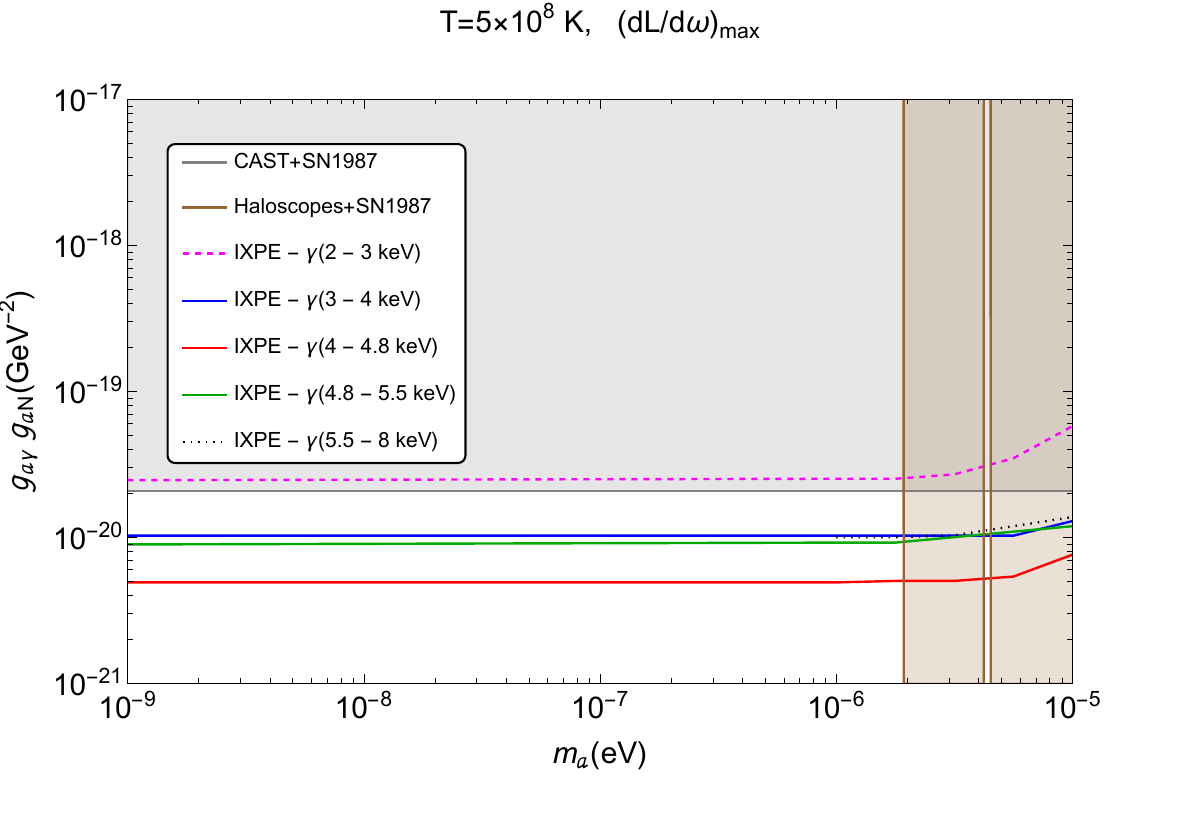}
\\
\includegraphics[width=0.49\textwidth]{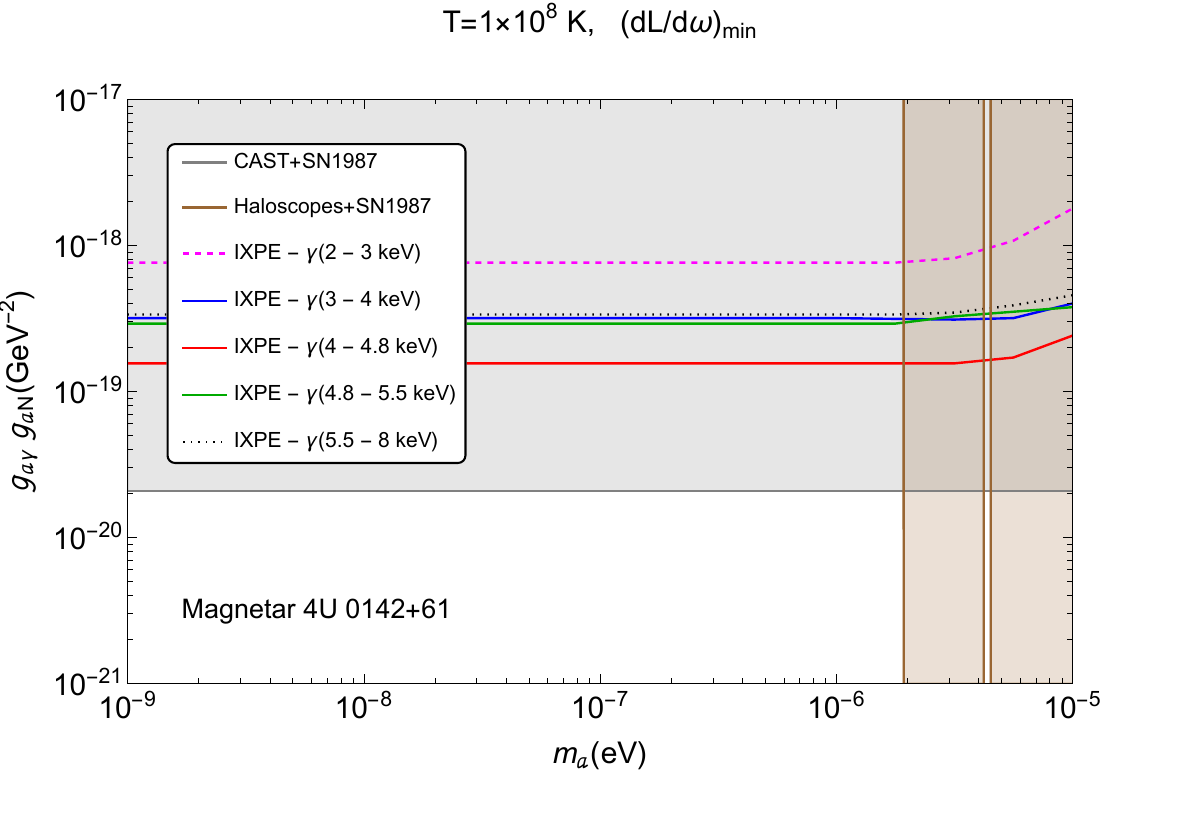}
\includegraphics[width=0.49\textwidth]{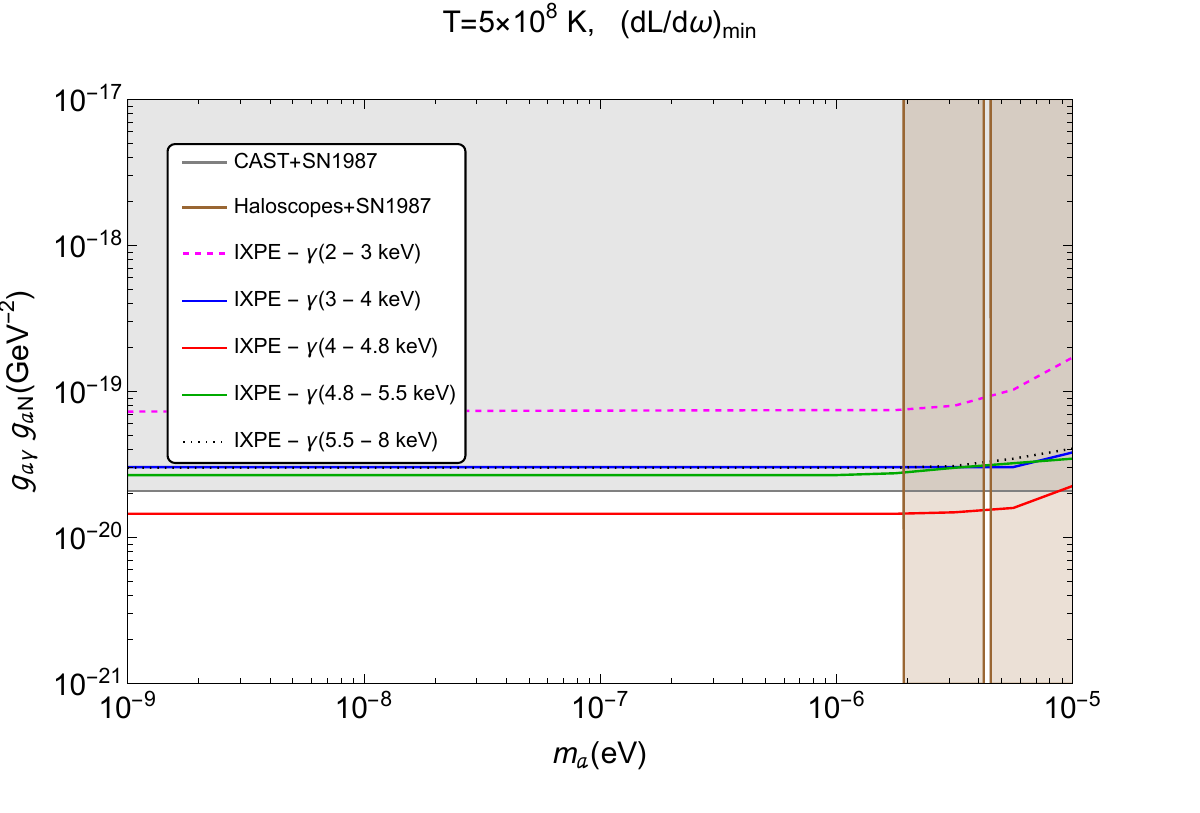}
  \caption{Bounds on the parameter space of ALP mass $m_a$~(eV) and the product of ALP-photon and ALP-nucleon couplings $g_{a\gamma} g_{a {\rm N}}$ (GeV$^{-2}$) for the magnetar 4U 0142+61. Results using a core temperature of $T^\infty = 1~(5) \times 10^8$~K are shown in the panels on the left (right) column. Results corresponding to the most optimistic and conservative luminosities (from varying the EoSs and magnetar masses) for a given core temperature are displayed on the top and bottom rows, respectively. The different colors in a given panel correspond to results from different energy bins. The horizontal shaded region is the combined constraint on the product of $g_{a\gamma}$ from CAST~\cite{CAST:2017uph} and $g_{aN}$ from SN1987A~\cite{Giannotti:2017hny}. The vertical shaded region is the haloscope constraint on $g_{a\gamma}$ from ADMX~\cite{ADMX:2021nhd, AxionLimits} combined with the $g_{aN}$ constraint from SN1987A.}
  \label{fig:const1}
\end{figure}

\begin{figure}
  \centering
\includegraphics[width=0.49\textwidth]{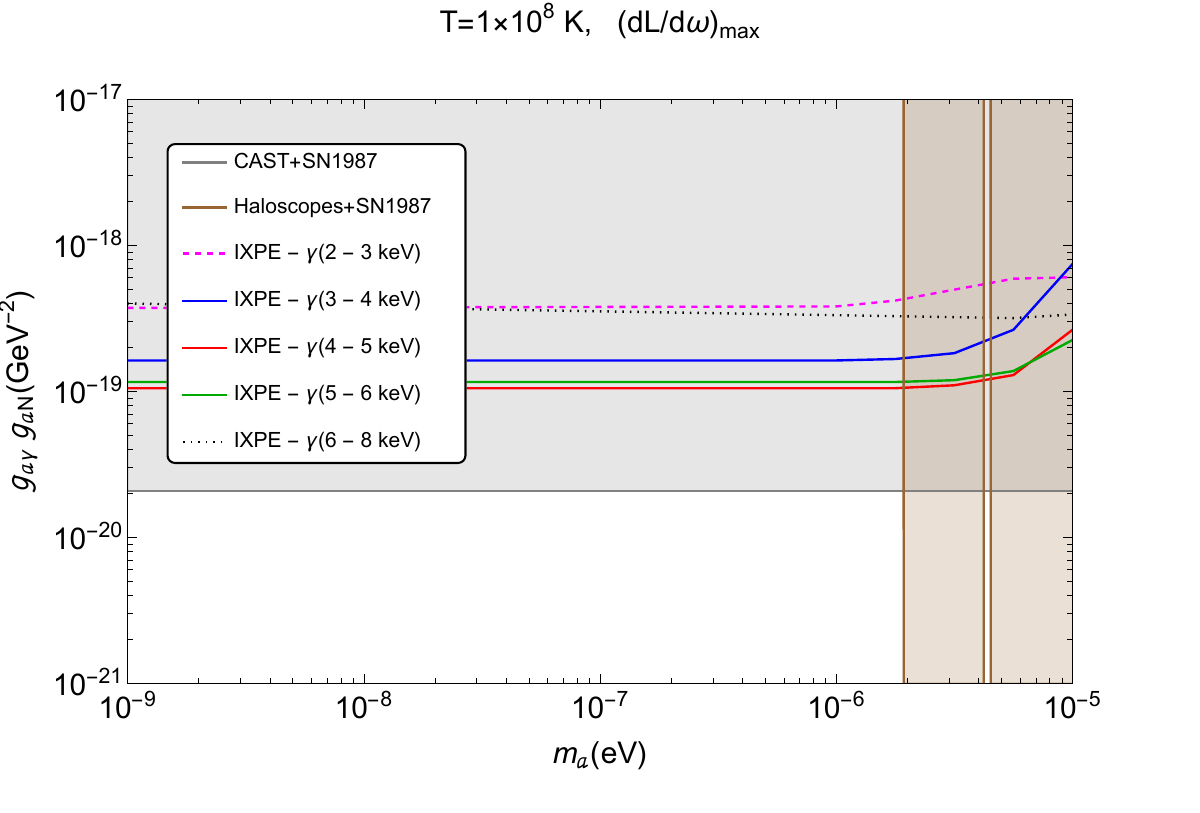}
\includegraphics[width=0.49\textwidth]{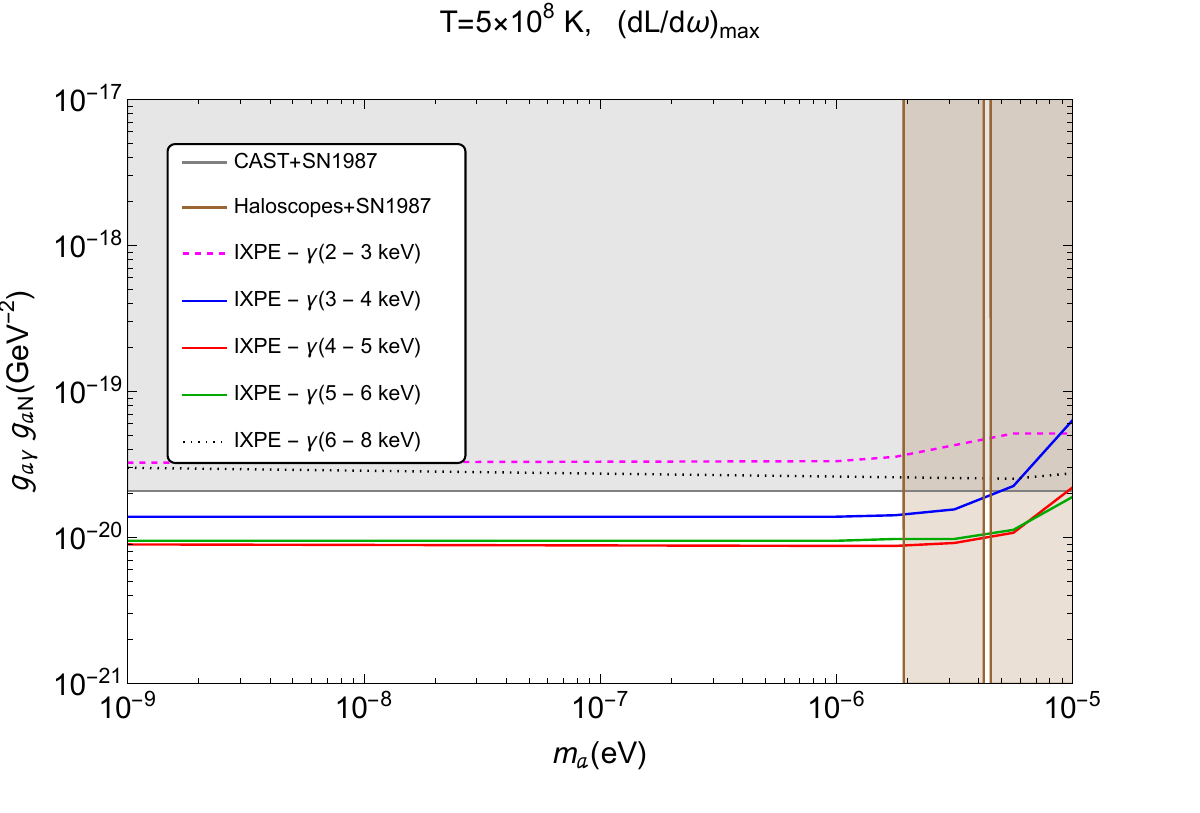}
\\
\includegraphics[width=0.49\textwidth]{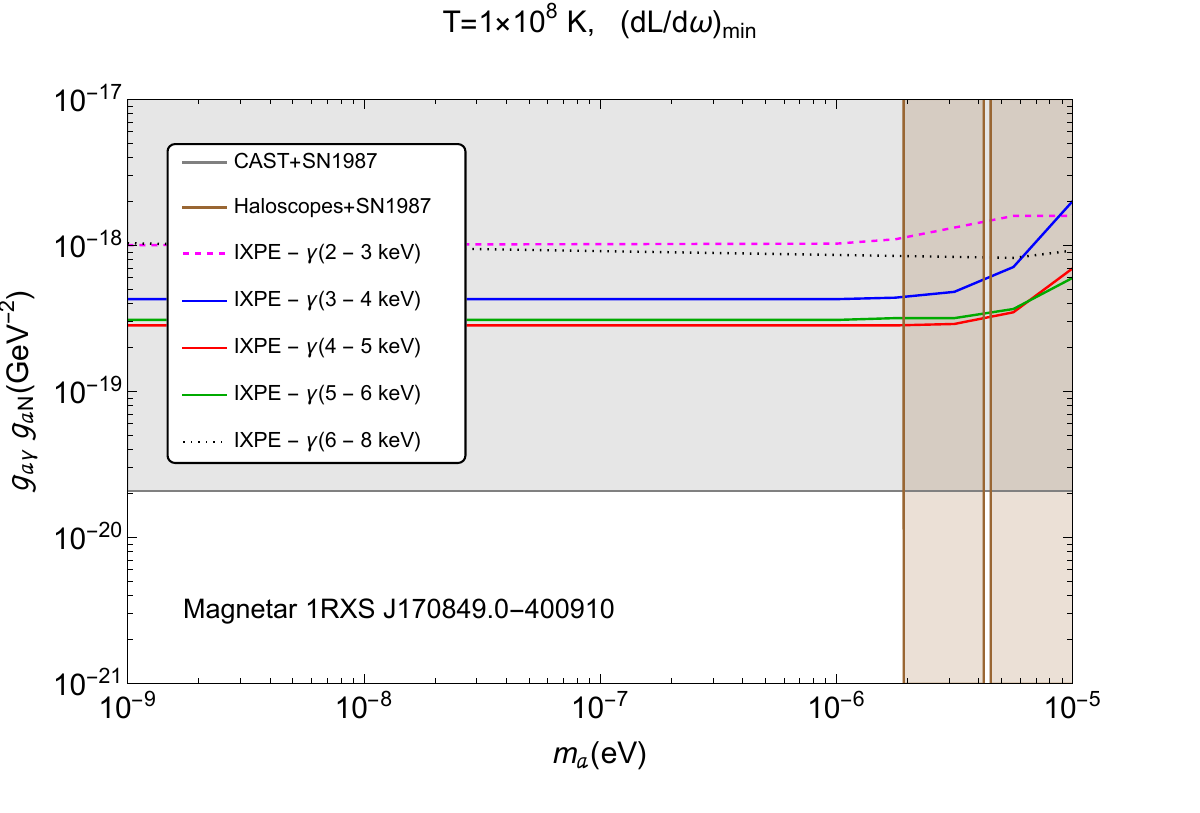}
\includegraphics[width=0.49\textwidth]{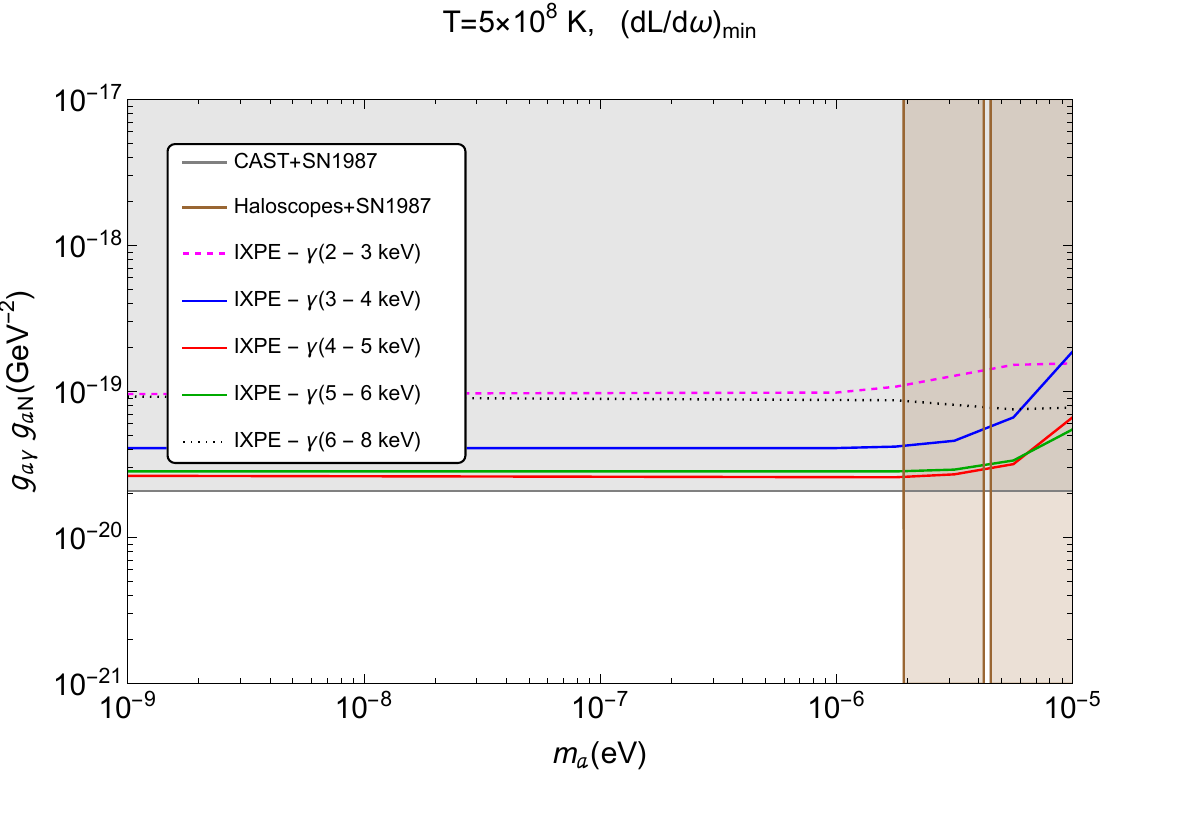}
  \caption{Bounds on the parameter space of ALP mass $m_a$~(eV) and the product of ALP-photon and ALP-nucleon couplings $g_{a\gamma} g_{a {\rm N}}$ (GeV$^{-2}$) for the magnetar 1RXS J170849.0-400910. Results using a core temperature of $T^\infty = 1~(5) \times 10^8$~K are shown in the panels on the left (right) column. Results corresponding to the most optimistic and conservative luminosities (from varying the EoSs and magnetar masses) for a given core temperature are displayed on the top and bottom rows, respectively. The different colors in a given panel correspond to results coming from different energy bins. The horizontal shaded region is the combined constraint on the product of $g_{a\gamma}$ from CAST~\cite{CAST:2017uph} and $g_{aN}$ from SN1987A~\cite{Giannotti:2017hny}. The vertical shaded region is the haloscope constraint on $g_{a\gamma}$ from ADMX~\cite{ADMX:2021nhd, AxionLimits} combined with the $g_{aN}$ constraint from SN1987A.}
  \label{fig:const2}
\end{figure}

The main principle of finding our constraints is as follows. We assume that, in the ideal case, the purely astrophysical modelling of the polarization degree and angle in each energy bin would provide a complete explanation for the {\it IXPE} observations (that is, the theoretically-predicted values using standard astrophysics would exactly match the experimentally-obtained values). Then, using the numerical version of Section~\ref{sec:poltheory}, which describes how Stokes $I$, $Q$, and $U$ would change in the presence of ALP-photon conversion, we calculate how much the (PD, PA) values would deviate from their central experimental values (as given in Tables~\ref{tab:mag1exp} and \ref{tab:mag2exp}), if such conversions were indeed to occur. We then restrict the possible ALP couplings to those that do not take the (PD, PA) from the original experimental values to a place on the plane outside the confidence contours (the ellipse outlines), shown in Figs.~\ref{fig:exp1} and~\ref{fig:exp2}, given for the two magnetars in Refs.~\cite{2022Sci...378..646T,2023ApJ...944L..27Z}. Finally, for both magnetars, we repeat this process four times: we use the upper and lower limits on the differential luminosity $dL_a^{\infty}/d\omega_a^{\infty}$ (obtained by varying the possible equations of state and masses), each in combination with reasonable lower ($T^{\infty}=1\times 10^8$~K) and higher ($T^{\infty}=5\times 10^8$~K) limits of magnetar core temperatures.

The specific steps of the constraint calculations are as follows. Using the numerical versions of Eq.~\eqref{key} through Eq.~\eqref{Eqpa}, we take as inputs different coupling products $g_{a\gamma}g_{aN}$ and different relative angles $\Delta \psi$ between the magnetar and detector frames, to produce points blanketing the polar plane of (PD, PA) within the bounds of the (elliptical) experimental error. As we increase the strength of the ALP couplings $g_{a\gamma}g_{aN}$, this causes the total ALP-inclusive (PD, PA) to move farther and farther from the original experimental value and towards the error contours displayed in Figs.~\ref{fig:exp1} and~\ref{fig:exp2}. The effect of the increase in coupling can be seen through the path traced out by (PD, PA) (for each relative angle) running between the red centerpoint and the blue boundary of the confidence region, for each relative angle. We restrict the range of allowable ALP-modified (PD, PA) values to those that stay within the 68\% (50\%) C.L. regions for the first (second) magnetar. 

Next, in each panel of Figs.~\ref{fig:constblob1} and~\ref{fig:constblob2}, the values for the allowed coupling products $g_{a\gamma}g_{aN}$ are plotted with respect to the possible relative orientations $\Delta \psi$ of the magnetars, with the different panels corresponding to different energy bins (that is, the different `ellipses' in Figs.~\ref{fig:exp1} and~\ref{fig:exp2}). 
Panels on the left column display the results assuming the lower magnetar core temperature $T^\infty=1\times 10^8$ K; those on the right column display the results with the higher temperature $T^\infty=5\times 10^8$ K. Within each panel, the brighter color corresponds to the results assuming the higher ALP luminosity bound; the fainter color corresponds to the results assuming the lower luminosity bound. In a given panel, each column of blue for a given $\Delta \psi$ corresponds to the path in the (PD, PA) polar plots that would be traced out between the experimental centerpoint and the edge of the confidence region by slowly increasing $g_{a\gamma}g_{aN}$. 
For these plots, we observe a somewhat `wave'-like pattern due to the fact that the experimental central value is not equidistant from all sides of the ellipse. Thus, the modification of the (PD, PA) due to ALPs for certain relative orientations will cause the ALP-inclusive (PD, PA) to reach the ellipse border much earlier (that is, at lower coupling products) than for other relative orientations. 

Finally, we take the most conservative (largest) choice of the coupling product $g_{aN}g_{a\gamma}$, out of all possible relative orientations, for each combination of energy bin, temperature, and luminosity, and plot them in Figs.~\ref{fig:const1} and~\ref{fig:const2} as a function of ALP mass $m_{a}$.\footnote{We note that Figs.~\ref{fig:constblob1} and~\ref{fig:constblob2} are calculated in the regime of small ALP mass $m_a$, and as such their results are valid for the constant parts of Figs.~\ref{fig:const1} and~\ref{fig:const2} before the `upturn' in the constraints. The full constraints as a function of mass in Figs.~\ref{fig:const1} and~\ref{fig:const2} are calculated without that assumption.} This yields our results on the plane of $g_{aN}g_{a\gamma}$ versus $m_{a}$. 
Each panel thus displays the constraints for the five different energy bins, for a given combination of core temperature and luminosity. These results are also plotted against the constraints obtained by considering the existing bounds on $g_{a\gamma}\lesssim 6.6\times 10^{-11}~{\rm GeV}^{-1}$ from CAST~\cite{CAST:2017uph} and on $g_{aN}\lesssim 3.2\times 10^{-10}~{\rm GeV}^{-1}$ from  SN1987A~\cite{Giannotti:2017hny}, with the parameter space ruled out by the combination of those observations shaded in grey (horizontal shaded region): $g_{a\gamma}g_{aN}\lesssim 2.1\times 10^{-20}~{\rm GeV}^{-2}$. In addition, we include the $g_{a\gamma}$ constraint from haloscope experiments like ADMX~\cite{ADMX:2021nhd, AxionLimits} which, in combination with the SN1987A constraint~\cite{Giannotti:2017hny}, rules out the parameter space for `higher' $m_a$ values (vertical shaded region). We see that our ALP constraints are competitive with (within an order of magnitude of) the existing CAST+SN1987A constraints.   

Note that we only show the most robust current constraints from CAST and SN1987A in Figs.~\ref{fig:const1} and~\ref{fig:const2}. Other constraints exist which are, in principle, stronger. For instance, a factor of 5 improvement over the SN1987A limit on $g^2_{aN}$ was obtained in Ref.~\cite{Beznogov:2018fda} using the hot young neutron star in the supernova remnant J1732, assuming that the thermal evolution of J1732 is dominated by neutrino cooling due to bremsstrahlung reactions involving unpaired neutrons in the core. Similarly, new astrophysical limits on $g_{a\gamma}$ from horizontal branch stars~\cite{Dolan:2022kul} and from pulsar data~\cite{Noordhuis:2022ljw} exist which are better than CAST, but are likewise subject to various astrophysical modeling uncertainties.

\section{Discussion and conclusions}
\label{sec:conclusion}

The bounds shown in Figs.~\ref{fig:const1} and~\ref{fig:const2} give the highest allowed values for the coupling product, considering all possible relative orientations, that do not cause the ALP-inclusive theoretical prediction to deviate from the experimental result by more than the confidence region around the central value of the data. This method of constraint is thus agnostic about the astrophysical model, insisting only that the correct astrophysical model will reproduce the central value of the experimental data; ALP contributions to the actual spectropolarimetric detections from {\it IXPE} are required to be sub-leading.

We first note that, just as both panels of Fig.~\ref{fig:ALP_dLdw} also imply, Figs.~\ref{fig:const1} and~\ref{fig:const2} demonstrate that the greatest uncertainty in our results comes from the different possible core temperatures. For a given core temperature, the differences in the constraints we get from using conservative versus optimistic selections of EoS and magnetar mass (the different `luminosities') are much smaller than the differences in constraints obtained from using different temperatures.

Continuing to compare the four graphs of Figs.~\ref{fig:const1} and~\ref{fig:const2} for each magnetar, we observe that the constraints obtained by simultaneously using the higher magnetar temperature ($5 \times 10^8$~K) and most optimistic luminosity selection correspond to the strongest bounds for both magnetars. This is expected: Eq.~\eqref{eq:Ia} shows how increasing the luminosity (which includes the temperature) will increase the $I_a$ (and thus $p_a$), for a given coupling product $g_{a\gamma}g_{aN}$. Thus, all else being equal, a smaller coupling product is needed for this highest-luminosity, higher-temperature case to reach the bounds of the confidence region, than that would be needed for the other three cases. This leads to smaller maximum coupling products in the `wave' shapes (Figs.~\ref{fig:constblob1} and~\ref{fig:constblob2}) and thus a tighter constraint on the coupling product, as shown in Figs.~\ref{fig:const1} and~\ref{fig:const2}. 

Examining the patterns within each subfigure of Figs.~\ref{fig:constblob1} and~\ref{fig:constblob2}, and the corresponding results in Figs.~\ref{fig:const1} and~\ref{fig:const2}, respectively, we notice that with all else held constant, the strongest ALP constraints are obtained in the middle energy bins of the {\it IXPE} 2~--~8~keV band: in the third (4.8~--~5.5~keV) bin for 4U 0142+61, and the third (4~--~5~keV) or fourth (5~--~6~keV) bins for 1RXS J170849.0-400910. A combination of two trends acting in opposite directions, one experimental and one theoretical, helps explain this optimization. 
On the observational side, as we go higher in energy, {\it IXPE} obtains fewer counts. This is due to the flux from both types of magnetar emission decreasing with increasing energy: the thermal/blackbody emission peaks below 2 keV and decays on both sides; the power-law RCS emission indeed follows a power-law distribution and decays exponentially with increasing energy~\cite{2022Sci...378..646T, 2023ApJ...944L..27Z}. Thus, with fewer counts at higher energies,\footnote{The effective area of the {\it IXPE} detectors, as displayed in Fig.~\ref{fig:detector}, also exhibits some energy dependence, especially in how much it drops off above 6 keV. This further contributes to the decrease in counts, and worse statistics, at higher energies.} the statistical uncertainties (as shown through the ellipse size for a given energy bin) will naturally increase. However, from theory, the effect of the ALP contribution parameter $p_a$ actually increases with energy, since the probability of conversion increases with energy in the energy band of our interest (from Eq.~\eqref{prob1} and Eq.~\eqref{prob2})~\cite{Fortin:2018ehg}, but mainly since the production from the core [cf. Eq.~\eqref{prod1} and Fig.~\ref{fig:ALP_dLdw}] increases with energy. The combination of these countervailing tendencies thus results in the fact that although lower (higher) energy bins have lower errors (higher ALP contribution), the other trend, acting in the opposite direction, decreases the strength of the constraints obtained at that energy. Thus, the best constraints, in the end, come from one of the bins in the middle of the energy band. 

\begin{figure}[t!]
  \centering
\includegraphics[width=0.6\textwidth]{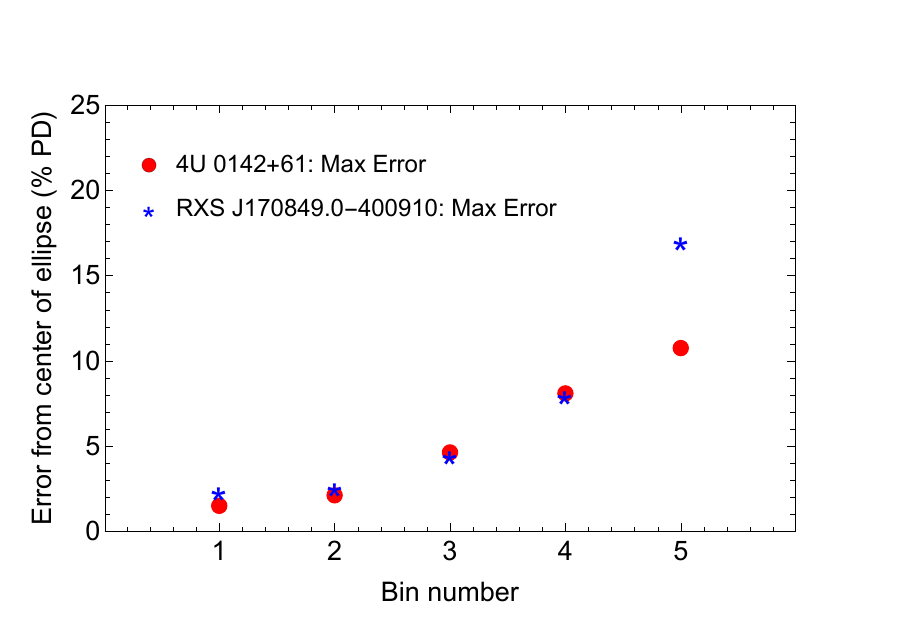}
\caption{The maximum error (in percentage) from the observational central data point to the edge of the ellipses (68\% and 50\% C.L. uncertainty contours) in the (PD, PA) plane in Figs.~\ref{fig:exp1} and~\ref{fig:exp2}, for magnetars 4U 0142+61 and RXS J170849.0-400910, respectively. The horizontal axis depicts the energy bin number in the sequence of five bins between 2~--~8~keV, with the specific bin widths included in Tables~\ref{tab:mag1exp} and \ref{tab:mag2exp}. 
}
  \label{fig:errorbin}
\end{figure}

We reiterate the main takeaway message about the observational status through Fig.~\ref{fig:errorbin}. Here, we plot the maximum\footnote{We take the maximum, as we are searching for the most conservative constraint on the coupling product.} distance from the central value of the data to the elliptical confidence level contours on the (PD, PA) polar plots for each energy bin and for each magnetar. It is clear that for bins at higher energies, the observational errors are much more significant than the bins at lower energies, for both magnetars. This effect counters the higher ALP contribution for the higher energy bins. Future improvements in those higher energy bins from the observational side are therefore critical for obtaining stronger ALP constraints.

Through future improvements on the astrophysical theoretical side (\textit{i.e.},\ with the modelling matching approaching ever more closely the centerpoint of the experimental results), we will increase confidence in our ALP constraints as the astrophysical theoretical models trend towards fully describing the observational results, which we already assume to be the case. 
Simultaneously, improved constraints would come through {\it IXPE} observing magnetars for even longer durations to improve statistics, especially at the higher energies, as well as through {\it IXPE} observing younger magnetars (those with higher luminosity and stronger magnetic fields, with the effect of the former clearly seen in the comparisons made by this paper) and located as close as possible to Earth. 
Finally, since the ALP luminosity spectrum from a source with temperature $T$ peaks at energies of a few times $T$ \cite{Raffelt:1996wa,Fortin:2021cog} (see also Fig.~3 in Ref.~\cite{Fortin:2021sst}), corresponding to several tens of keV for the two magnetars studied here, a hard $X$-ray polarization mission such as {\it XL-Calibur}~\cite{2021APh...12602529A, Krawczynski:2019ofl}, designed for the 15~--~80~keV range, would more than ideally complement the capabilities of {\it IXPE} through enabling a study with ALP couplings stronger than any of those in the energy range examined in this current work.

\acknowledgments
E.G. acknowledges many helpful discussions with other members of the {\it IXPE} collaboration, especially Roberto Turolla, Roberto Taverna, Michela Negro, and Fabio Muleri, as well as with Takuya Okawa. K.S. would also like to thank Matthew Baring for many illuminating discussions. The work of B.D. was partly supported by the U.S. Department of Energy under grant No. DE-SC 0017987. F.H. and K.S. are supported in part by DOE grant DE-SC0009956. K.S. would like to thank the Simons Center for Geometry and Physics for hospitality, where part of this work was done. S.P.H. is supported by the U.~S. Department of Energy grant DE-FG02-00ER41132 and the NSF grant PHY 21-16686. J.F.F. is supported by NSERC. E.G. and H.K. acknowledge NASA support through the grants NNX16AC42G, 80NSSC20K0329, 80NSSC20K0540, NAS803060, 80NSSC21K1817, 80NSSC22K1291, and 80NSSC22K1883. 

\appendix

\vspace{1cm}

%%%%%%%%%%%%%%%%%%%%%%%%%%%%%bibliography%%%%%%%%%%%%%%%%%%%%%%%%

\bibliography{draft}
\bibliographystyle{JHEP}

\end{document}